# Consumer Segmentation and Participation Drivers in Community-Supported Agriculture: A Choice Experiment and Partial Least Squares Structural Equation Modelling Approach

Sota Takagi[1*], Miki Saijo[1], Takumi Ohashi[1,2]
[1]Institute of Science Tokyo, 2-12-1 Ookayama Meguro-ku, Tokyo 152-8550, Japan
* takagi.s.5b1d@m.isct.ac.jp
[2]Chulalongkorn University, Phayathai Road, Pathumwan, Bangkok 10330, Thailand

Abstract
As the global food system faces increasing challenges from sustainability, climate change, and food security issues, alternative food networks like Community-Supported Agriculture (CSA) play an essential role in fostering stronger connections between consumers and producers. However, understanding consumer engagement with CSA is fragmented, particularly in Japan where CSA participation is still emerging. This study aims to identify potential CSA participants in Japan and validate existing theories on CSA participation through a quantitative analysis of 2,484 Japanese consumers.
Using choice experiments, Latent Class Analysis, and Partial Least Squares Structural Equation Modeling, we identified five distinct consumer segments. The "Sustainable Food Seekers" group showed the highest positive utility for CSA, driven primarily by "Food Education and Learning Opportunities" and "Contribution to Environmental and Social Issues." These factors were consistently significant across all segments, suggesting that many Japanese consumers value CSA for its educational and environmental benefits. In contrast, factors related to "Variety of Ingredients" were less influential in determining participation intentions. The findings suggest that promoting CSA in Japan may be most effective by emphasizing its role in environmental and social impact, rather than focusing solely on product attributes like organic certification, which is readily available in supermarkets. This reflects a key distinction between CSA adoption in Japan and in other cultural contexts, where access to organic produce is a primary driver.
This research provides practical insights for CSA organizations, policymakers, and producers, offering recommendations on how to better promote CSA participation in Japan and contribute to building a sustainable food system.

## 1. Introduction

The industrialization and globalization of the food market have necessitated a transition from conventional food supply systems to sustainable alternatives in response to escalating environmental concerns. Short food supply chains (SFSCs) have gained prominence for their role in enhancing sustainability, with Community-Supported Agriculture (CSA) emerging as a key model that fosters direct producer-consumer relationships (Princen 1997; Galli and Brunori 2013; Abbate et al. 2023). Unlike conventional distribution systems that rely on intermediaries, the CSA model allows consumers to prepay for seasonal produce, providing farmers with financial stability while promoting local food systems (Worden 2004). CSA also offers consumers opportunities to engage in farming activities, fostering deeper connections with food production and sustainability (Blättel-Mink et al. 2017; Egli, Rüschhoff, and Priess 2023).

Despite the increasing global interest in CSA as a sustainable food system, research on consumer participation remains fragmented, particularly in Japan, where CSA adoption is still limited. While previous studies have explored CSA participation factors in various contexts, no research has quantitatively validated CSA participation models in Japan (Takagi et al. 2024). Furthermore, existing studies have not fully examined how socio-cultural and psychological factors influence CSA participation, nor have they accounted for consumer heterogeneity.

To address these gaps, this study aims to identify potential CSA consumers in Japan and quantitatively examine the factors influencing their participation by validating CSA participation models. By clarifying the determinants that shape consumer engagement in CSA, this research contributes both theoretical and practical insights that can support the promotion of CSA and the development of sustainable food systems in Japan. This study makes three key contributions.

First, it segments consumers based on their preferences and motivations, allowing for a deeper understanding of participation heterogeneity. Second, it provides the first quantitative validation of CSA participation models in Japan, offering empirical evidence to refine theoretical frameworks on CSA engagement. Third, it extends CSA research beyond Western contexts by analyzing socio-cultural influences on CSA adoption in Japan, contributing to a broader cross-cultural perspective on sustainable food systems.

## 2. Literature review

### 2.1 Consumer motivations and segmentation in CSA participation

CSA has gained attention as an alternative food distribution model that fosters direct relationships between farmers and consumers. However, despite its potential benefits, research on CSA adoption in Japan remains limited. While CSA participation models have been explored in various contexts, prior studies have primarily examined psychological motivations and the perceived advantages and disadvantages of CSA participation (Datai et al. 2023). These studies have not quantitatively measured the relative impact of these factors on consumer participation intentions, leaving a gap in understanding the determinants of CSA engagement.

Existing literature on sustainable food consumption highlights the role of socio-economic, psychological, and behavioral factors in shaping consumer decisions (Fujisaki et al. 2025). Studies conducted in Western contexts have identified key consumer segments, such as environmentally motivated consumers, health-conscious consumers, and price-sensitive consumers (Zepeda and Li 2006; Vermeir and Verbeke 2008). However, similar segmentation research in Japan is scarce, despite evidence suggesting significant consumer heterogeneity in CSA adoption (Toriyama, Sato, and Suzuki 2021). Some Japanese consumers prioritize food safety and organic produce, while others are motivated by community engagement and sustainability values. Without a structured understanding of these segments,

CSA organizations in Japan lack the necessary insights to tailor their outreach strategies effectively.

## 2.2 Trust and consumer perceptions in CSA participation

Trust and perception play a crucial role in shaping consumer behavior, particularly in CSA participation, which requires an upfront financial commitment. Previous studies have linked trust in organic agriculture to environmentally conscious purchase intentions and sustainable food consumption behaviors (Lazaroiu et al. 2019). Since CSA involves a long-term relationship between consumers and farmers, trust in food production is essential for engagement. Research on sustainable brand management suggests that consumer loyalty is shaped by perceived value, emotional associations, and confidence in product quality (Majerova et al. 2020). In addition to sustainability concerns, consumer trust is also linked to the perceived benefits of fresh food consumption. A study on Generation Z university students in Romania identified various clusters of young consumers based on their fruit and vegetable consumption habits. The study found that while many individuals recognize the importance of fresh food for maintaining health and well-being, some consumer groups do not prioritize fresh produce in their diet (Pocol et al. 2021). While these studies provide valuable insights into consumer trust in organic food and fresh produce, they do not comprehensively address consumer behavior in the context of CSA participation. Most existing research focuses on trust in food quality, freshness, and production transparency (J. Chen et al. 2019; Hanson, Concepcion, and Volpe 2024), but there is no research that has integrated and quantitatively examined these factors in the context of CSA participation. Because CSA requires both a financial commitment and a long-term relationship between consumers and producers, more research is needed to understand how trust influences consumer decisions to participate in this unique food system.

## 2.3 Cultural and regional differences in CSA participation

Cultural factors significantly influence CSA participation decisions. Comparative studies between the U.S. and France have shown variations in consumers' willingness to invest time and effort in CSA participation as a means of verifying food origin (Peterson, Taylor, and Baudouin 2015). Furthermore, a choice experiment conducted in Connecticut found that risk mitigation strategies for poor harvests significantly increased consumers' likelihood of participating in CSA (Yu et al. 2019). These findings suggest that motivations and barriers to CSA adoption may differ across regions, necessitating localized investigations into the factors shaping CSA participation.

## 2.4 Research gaps

While previous studies have explored the motivations for CSA participation, few have quantified the impact of socio-cultural and psychological factors on participation intentions, particularly in Japan. Additionally, existing research has not systematically classified consumer heterogeneity in CSA adoption, making it difficult to tailor effective engagement strategies (Toriyama, Sato, and Suzuki 2021). Cultural differences also shape participation behaviors. However, these dynamics remain underexplored in the Japanese context. Based on these gaps, this study addresses the following research questions:

RQ1: Who are the potential CSA participants in Japan, and how can they be classified into distinct consumer segments?

RQ2: What are the key factors influencing consumer intentions to participate in CSA?

RQ3: How do the key factors influencing consumer intentions to participate in CSA differ between consumer segments?

## 3. Materials and methods

### 3.1 Study design

To investigate Japanese consumers' preferences for CSA-distributed foods and their intentions to participate in CSA programs, we conducted an online questionnaire survey. Based on previous research indicating that farmers' market participants are potential CSA consumers
(Pisarn, Kim, and Yang 2020), we included questions related to farmers' markets in our survey.

The questionnaire used in this study consisted of three main components: (1) a choice experiment involving the selection of baskets of domestically grown vegetables, (2) questions related to the CSA participation model, and (3) questions about CSA participation intentions and awareness.

Based on the choice experiment data, we conducted a Latent Class Analysis (LCA) to identify distinct consumer segments associated with preferences for CSA attributes (RQ1). We then applied a conditional logit model to elucidate the utility of CSA attributes in product choice and estimated the marginal willingness to pay (MWTP) for each identified segment. To uncover the factor structure of the CSA participation model (RQ2), we performed an Exploratory Factor Analysis (EFA). The derived factors were then incorporated into a Partial Least Squares Structural Equation Modeling (PLS-SEM) to analyze the relationships between the socio-cultural environment, perceived gains, and the intention to participate in CSA. Furthermore, to examine how each factors shape participation intentions across different consumer segments (RQ3), PLS-SEM was applied both to the entire sample and to individual segments, allowing for a comparative analysis of consumer decision-making processes across different groups.

### 3.2 Data collection and sample

This study utilized an internet survey targeting Japanese consumers, conducted through a market research company. Initially, the lead author developed an online questionnaire and conducted a preliminary survey with 46 participants, recruited through convenience sampling, to ensure the clarity of the choice experiment content and the reliability of the question items based on previous research. The main survey was then conducted anonymously through a Japanese monitoring company from March 1 to 2, 2024. Participants were informed of the study's purpose and the potential for the results to be published in academic papers through an introductory face sheet. Electronic consent was obtained from all participants prior to the commencement of the survey.

To ensure that participants had an active role in food-related decision-making at home, the target population consisted of individuals who responded “Always” or “Sometimes” to the questions “Do you purchase ingredients yourself when eating at home?” and “Do you cook for yourself when eating at home?” The initial sample size was 3,020 participants. To mitigate concerns regarding careless responses in online surveys, several reverse-coded questions were incorporated. These reverse-coded items functioned as a screening mechanism to identify and exclude inconsistent or inattentive responses (Churchill 1979; Paulhus 1991; Swain, Weathers, and Niedrich 2008). After applying these screening criteria, 2484 valid responses were analyzed. The socio-demographic composition of the respondents is summarized in Table 1.

Table 1. Sociodemographic characteristics of respondents ($N$ = 2484).

| | | *N* | % |
|---|---|---|---|
| Gender | Male | 1020 | 41.1 |
| | Female | 1464 | 58.9 |
| Age | Average | 50.7 | |
| | 20-29 | 151 | 6.1 |
| | 30-39 | 426 | 17.1 |
| | 40-49 | 577 | 23.2 |
| | 50-59 | 633 | 25.5 |
| | 60+ | 697 | 28.1 |
| Region of residence | Hokkaido | 129 | 5.2 |
| | Tohoku | 119 | 4.8 |
| | Kanto | 943 | 38.0 |
| | Chubu | 385 | 15.5 |
| | Kinki | 485 | 19.5 |
| | Chugoku & Shikoku | 203 | 8.2 |
| | Kyusyu & Okinawa | 220 | 8.9 |
| Annual household income | Under 4 million JPY | 942 | 37.9 |
| | 4 to 6 million JPY | 584 | 23.5 |
| | Over 6 million JPY | 958 | 38.6 |
| Annual personal income | Under 4 million JPY | 1816 | 73.1 |
| | 4 to 6 million JPY | 344 | 13.8 |
| | Over 6 million JPY | 324 | 13.0 |
| Married | Yes | 1497 | 60.3 |
| | No | 987 | 39.7 |
| Household size | Single | 643 | 25.9 |
| | 2 | 781 | 31.4 |
| | 3 | 536 | 21.6 |
| | 4 | 394 | 15.9 |
| | 5 | 92 | 3.7 |
| | 6 or more | 38 | 1.5 |
| Under 18 in household | Yes | 668 | 26.9 |
| | No | 1816 | 73.1 |
| Full-time homemaker | Yes | 435 | 17.5 |
| | No | 2049 | 82.5 |
| Education | Primary and Secondary | 811 | 32.7 |
| | Vocational | 653 | 26.3 |
| | Undergraduate | 931 | 37.5 |
| | Postgraduate | 88 | 3.5 |

### 3.3 Questionnaire design

#### 3.3.1 Questionnaire structure

The questionnaire used in this study consisted of three sections:

1. Choice Experiment: Participants selected baskets of domestically grown vegetables to calculate the utility and marginal willingness to pay for each attribute.
2. CSA Participation Model: This section included 33 questions on the socio-cultural environment and 20 questions on the perceived balance of gains and losses, each using a 7-point Likert scale ranging from "1. Strongly Disagree" to "7. Strongly Agree."
3. CSA Participation Intentions and Awareness: Respondents' awareness of CSA and their willingness to participate were assessed. Awareness was measured through a four-category question, and intention to participate was measured using a 7-point Likert scale.

#### 3.3.2 Choice experiment

A choice experiment was used to examine consumer preferences and willingness to pay for CSA. Choice experiments are particularly effective in exploring preferences for products and attributes that are not widely available in actual market environments (Louviere, Hensher, and Swait 2000). The experiment included four attributes:

- Basket purchased from (CSA, farmers' market, supermarket),
- Organic certification label (present or absent),
- Pickup point from home (home, 10, 20, or 30 minutes by car or bus), and
- Price (2,000 yen, 3,000 yen, 4,000 yen, and 5,000 yen).

An orthogonal array was used to generate 16 choice sets for each product. Each choice set consisted of two options and an opt-out option. Respondents were asked to select their preferred option from each set, with the order of presentation randomized to minimize order effects.

Table 2 Attribute levels for choice experiment

| | Basket purchased from | Label | Pickup point | Price (JPY/basket) |
|---|---|---|---|---|
| Level 1 | CSA | Organic certified label | Delivered to home | 2000 |
| Level 2 | Farmers' market | No label | 10 minutes by car or bus | 3000 |
| Level 3 | Supermarket | | 20 minutes by car or bus | 4000 |
| Level 4 | | | 30 minutes by car or bus | 5000 |

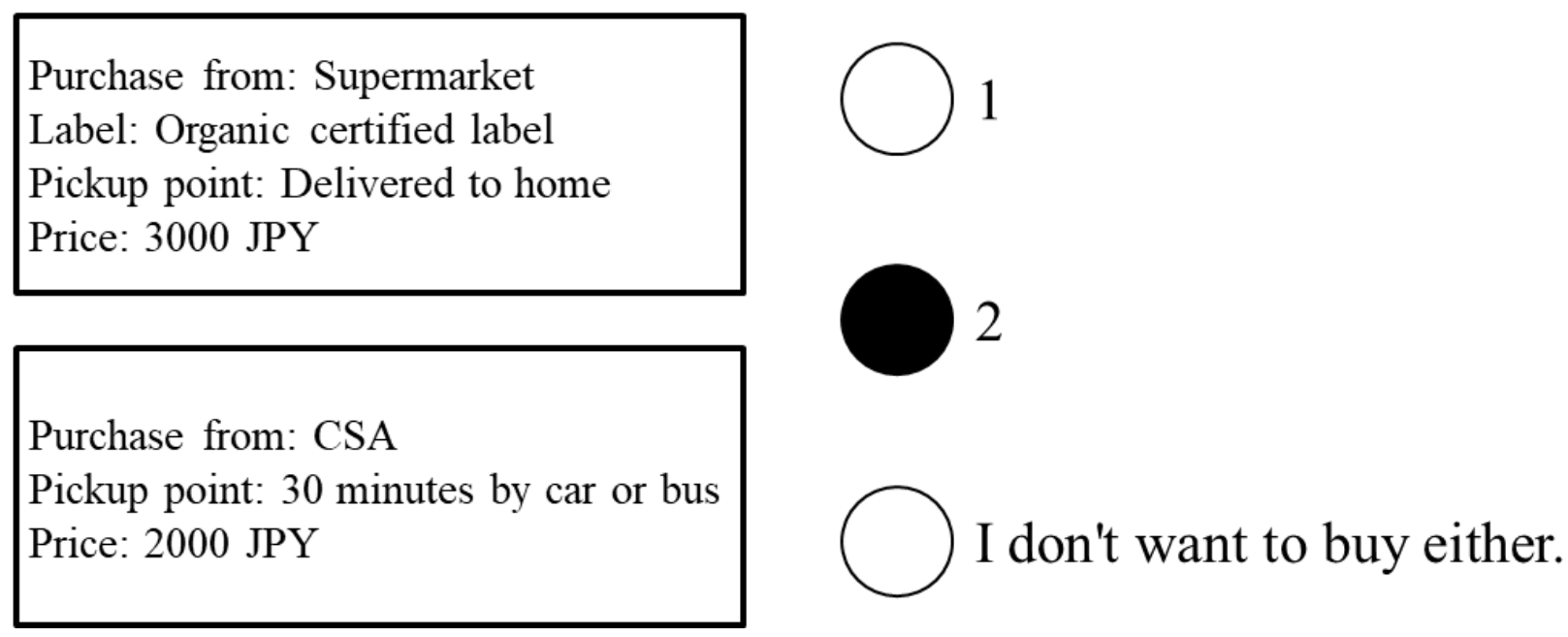

Fig. 1. Experiment screen presented to the respondents.

### 3.3.3 CSA participation model

This study examined the factors that influence Japanese consumers' intention to participate in CSA using the CSA participation model. This model indicates that consumers decide to participate based on a balance of expected gains and losses associated with CSA involvement, with this decision-making process shaped by the socio-cultural environment in which the consumer is embedded (Takagi et al. 2024).

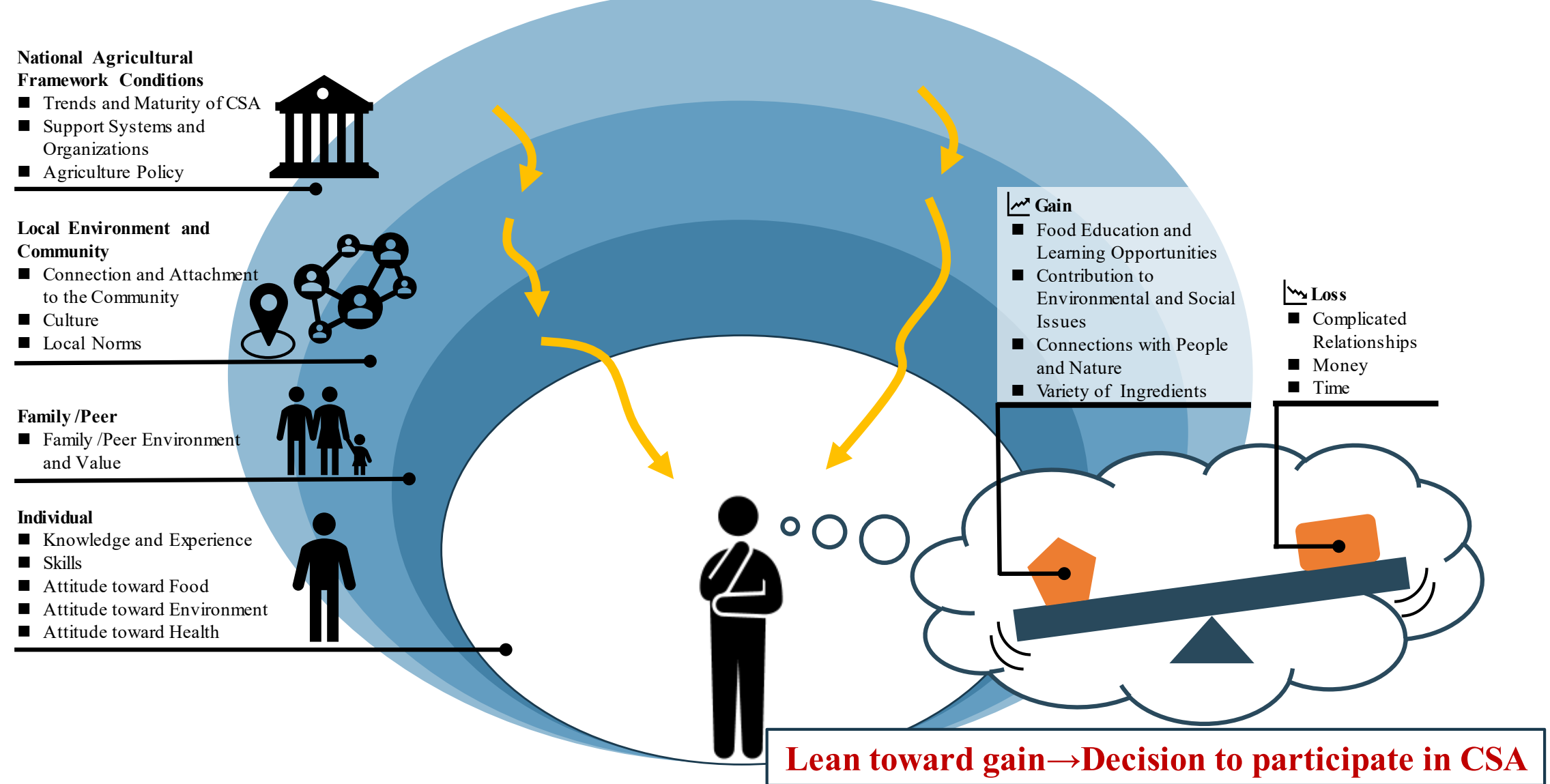

Fig. 2 CSA participation model. (Adopted and modified from (Takagi et al. 2024))

The socio-cultural environment was conceptualized across three levels: "Individual," "Family/Peer," and "Local Environment and Community." Gains included "Food Education and Learning Opportunities," "Contribution to Environmental and Social Issues," "Connections with People and Nature," and "Variety of Ingredients." Losses included "Complicated Relationships," "Money," and "Time." Given the niche status of CSA in Japan, the "National Agricultural Framework Conditions" level was excluded from the analysis.

The survey questionnaire was developed based on prior literature and adapted to align with the CSA participation model. Details of the questionnaire, including the references, are provided in Table A1 and A2. Since the primary objective of this study is to quantitatively validate the CSA participation model rather than to derive hypotheses directly from prior studies, the relationships examined in this study were not explicitly formulated as hypotheses based on previous literature but were instead structured to test the model's applicability in the Japanese context.

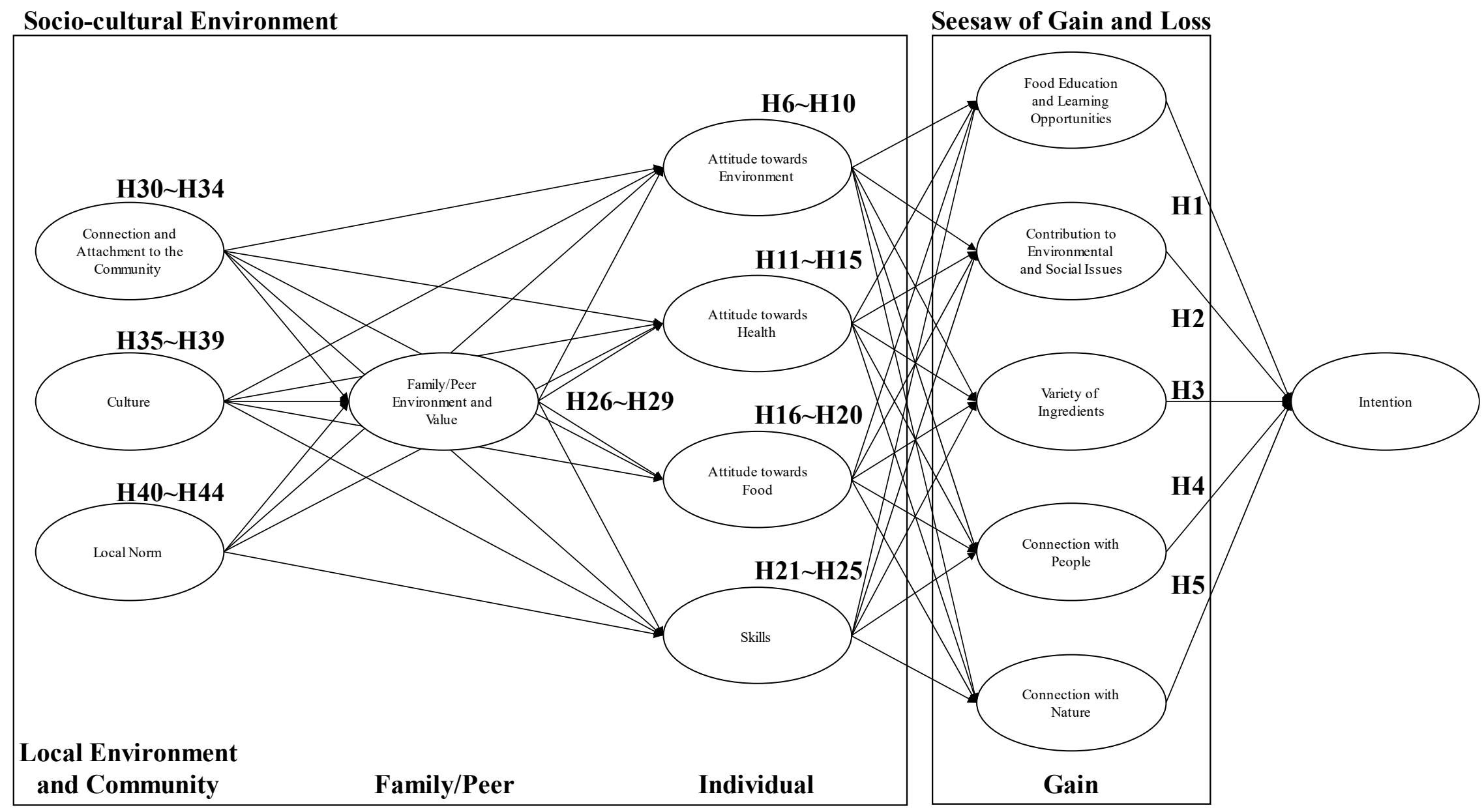

Fig. 3 Proposed conceptual model with hypotheses. (Hypotheses are listed in Table A3)

## 3.4 Data analysis

### 3.4.1 Latent class analysis

LCA was applied to the choice experiment response patterns to identify consumer segments within the sample. LCA assumes that the population consists of a finite number of latent classes and estimates the probability of each respondent belonging to each segment. LCA is suitable for identifying potential consumer groups that are likely to participate in CSA, which is a key objective of this study. The number of latent classes was determined based on the Akaike Information Criterion (AIC) and Bayesian Information Criterion (BIC), while also considering the interpretability of the obtained segments. In this study, the poLCA package (Linzer and Lewis 2011) in R version 4.1.2 (R Core Team 2021) was used for the LCA.

### 3.4.2 Conditional logit model

For each segment identified by latent class analysis, a conditional logit model was estimated to quantify the influence of each attribute on basket selection. The conditional logit model is a method that models the utility derived from individual choices and assesses the relative attractiveness of different options. In this model, consumers are assumed to define the utility provided by each attribute level as a linear function and select the alternative with the highest utility, thereby quantifying the impact of each attribute on product selection. Subsequently, the MWTP for each attribute level was calculated, reflecting the economic value consumers assign to different attribute levels. In this study, the clogit function from the survival package (Therneau et al. 2021) was used. The use of conditional logit models facilitates understanding of the determinants of consumer choice for each product attribute (McFadden 1972).

## 3.5 Reliability and validity analysis

### 3.5.1 Exploratory factor analysis

An EFA was conducted to determine the factor structure of the questionnaire items. This analysis included 32 questions related to the socio-cultural environment and 19 questions related to gains, excluding screening items, and was performed on a sample of 2,484 respondents. EFA was conducted separately for the socio-cultural environment and benefits. The number of extracted factors was determined using principal axis factoring (PAF) and parallel analysis (Horn 1965), utilizing SPSS syntax provided by (O'connor 2000). EFA was performed using principal axis factoring with direct oblimin rotation, applying a factor loading cutoff of 0.3. Observed variables with cross-loadings (i.e., loading on two or more factors) were iteratively removed to achieve a simple structure.

Internal consistency was assessed using Cronbach's alpha, with a threshold of 0.7 for acceptable reliability. The resulting factor structure was then used to guide subsequent PLS-SEM analysis. Factor loadings and EFA results are presented in the results section.

### 3.5.2 Partial Least Square Structural Equation Modeling (PLS-SEM)

PLS-SEM was adopted to examine relationships between latent variables and validate the hypothesized model. The PLS-SEM model was constructed using survey results on intention to participate in CSA. PLS-SEM is a variance-based structural equation modeling technique suitable for exploratory research and models with complex relationships (Sarstedt, Ringle, and Hair 2017; J. Hair and Alamer 2022). This method allows for simultaneous estimation of multiple dependent relationships while accounting for measurement error in latent constructs.

The analysis was conducted using SmartPLS 4 (ver.4.1.0.6) software (Ringle et al. 2023). Reliability and validity of the measurement model were assessed by examining the outer loadings of each construct (threshold: 0.7), composite reliability (rho_a, rho_c; threshold: 0.7), AVE (threshold: 0.5). In addition, the variance inflation factor (VIF; threshold: 0.5) was used to evaluate the problem of multicollinearity, and the Fornell & Larcker criterion was used to evaluate discriminant validity.

A two-step process was used to create a model with "Attitude towards Environmental Activities" and "Environmentally-friendly Choices" factors as components of "Attitude towards Environment" (Sarstedt et al. 2019). The latent variable scores for "Attitude towards Environmental Activities" and "Environmentally-friendly Choices" were calculated using values obtained from executing a bootstrap procedure with 10000 resamples for each segment.

To test the significance of path coefficients, a bootstrap procedure with 10000 resamples was executed. This non-parametric approach provides robust estimates of standard errors and confidence intervals for model parameters, enabling statistical inference in PLS-SEM (Sarstedt, Ringle, and Hair 2017). The results of the PLS-SEM analysis reveal factors related to Japanese consumers' willingness to participate in CSA, providing practical insights for proposing appropriate intervention strategies.

## 4. Results

### 4.1 Consumer segmentation

LCA was conducted to segment respondents based on their preferences for basket attributes. After comparing AIC and BIC values while considering interpretability, the optimal number of segments was determined to be five. AIC and BIC values are provided in the Table A4. Based on the choice preferences of respondents in each segment, they were named "Conventional Shoppers," "Organic Enthusiasts," "Farmers' Market Advocates," "Home Delivery Preferers," and "Sustainable Food Seekers." The results are shown in Table 4. Their characteristics are as follows:

Conventional Shoppers (18.5%): This segment shows negative reactions to CSA and farmers' market options and shows a preference for conventional non-organic products. These consumers are sensitive to the distance they must travel to receive purchased goods, reacting negatively to greater distances from home. They are also most price-sensitive segment, suggesting a preference for conventional supermarkets and prioritizing proximity and low prices over other attributes.

Organic Enthusiasts (33.8%): This segment, the largest proportion, is the only one showing willingness to pay for organic certification labels. They are sensitive to pickup point distance, strongly disliking pickup locations far from home, similar to Conventional Shoppers. Overall, these consumers prioritize organic products and prefer convenient pickup locations.

Farmers' Market Advocates (11.3%): These segment shows a notable positive reaction to farmers' markets. They are slightly negative towards organic labeling, suggesting a preference for non-organic products or indifference to organic certification. This segment is sensitive to both price and travel distance for pickup point, preferring closer locations. They are likely to choose farmers' markets, especially those near their homes, and may be strongly influenced by price.

Home Delivery Preferers (10.0%): Similar to Conventional Shoppers, this segment prefers conventional supermarkets and prioritizes proximity and low prices. However, they are particularly negative towards CSA, farmers' markets, and long distances from home. While they are least sensitive to price compared to other segments, their main concern is convenience of receipt. These consumers likely prioritize the ease of having groceries delivered to their doorstep over other factors.

Sustainable Food Seekers (26.4%): This group is favorable towards both CSA and farmers' market, with a slight preference for farmers' markets. They are somewhat negative towards organic labels. They tend to prefer pickup point 10 minutes from home rather than home delivery. Their price sensitivity is relatively low, only slightly higher than the Home Delivery Preferers. This suggests that these consumers appreciate the benefits of both CSA and farmers' market.

Table 4. Marginal willingness to pay for attribute levels

| | | All | Conventional Shoppers | Organic Enthusiasts | Farmers’ Market Advocates | Home Delivery Preferers | Sustainable Food Seekers |
|---|---|---|---|---|---|---|---|
| Attribute | Level | *N* = 2484 | *N* = 460, 18.5% | *N* = 839, 33.8% | *N* = 280, 11.3% | *N* = 250, 10.0% | *N* = 655, 26.4% |
| Basket purchased from | ref. Supermarket | | | | | | |
| | CSA | -176.28*** | -228.11*** | -547.22*** | 92.26 | -2058.96*** | 286.21*** |
| | Farmers’ Market | -67.86 | -107.16* | -593.70*** | 260.72*** | -1551.70*** | 415.39*** |
| Label | ref. No label | | | | | | |
| | Organic | -65.32* | -110.57 | 449.78** | -60.51 | -308.17 | -98.55* |
| Pickup | ref. Delivered to home | | | | | | |
| | 10 minutes from home | -153.24*** | -78.98* | -505.43*** | -112.48* | -2683.93*** | 158.02* |
| | 20 minutes from home | -687.45*** | -712.84*** | -436.89* | -442.58*** | -3616.09*** | -465.90*** |
| | 30 minutes from home | -747.31*** | -695.97*** | -403.10* | -923.97*** | -4069.37*** | -812.24*** |

Unit is Japanese yen

Significance codes: 0 < *** < 0.001 < ** < 0.01 < * < 0.05

Willingness to pay: $WTP_{k|s} = -\left(\frac{\beta_{k|s}}{\beta_{price|s}}\right)$

## 4.2 Determinants of participate intention

### 4.2.1 Exploratory factor analysis

Parallel analysis was conducted to determine the number of factors to extract. The eigenvalues from our dataset were compared with the average eigenvalues from 1000 randomly generated datasets. Factors with eigenvalues larger than those from the random datasets were retained (Table A5). This analysis suggested that 9 factors for the socio-cultural environment and 5 factors for gains were appropriate. Subsequently, EFA was performed with the number of factors fixed to those obtained from the parallel analysis. After removing observed variables with cross-loadings and factor loadings below 0.3, it was found that no observed variables corresponded to one factor in the socio-cultural environment. Therefore, the number of factors was fixed at 8, and the analysis was repeated under the same conditions. As a result, the socio-cultural environment ultimately had 8 factors, and benefits had 5 factors, which were labeled as follows. While the gains aligned with the latent factors in the hypothetical model (Fig. 3), in the socio-cultural environment, "Attitude towards Food" was excluded, and "Attitude towards Environment" was split into "Attitude towards Environmental Activities" and "Environmentally-friendly Choices." To ensure the reliability of the measurement scales, Cronbach's alpha was iteratively evaluating whether the overall alpha value increased when any item was deleted. After this refinement process, the final Cronbach's alpha values for each construct are fixed. The results are shown in Tables 5 and 6.

Table 5 Pattern matrix of exploratory factor analysis for Socio cultural environment

| ***Factor (Cronbach's alpha)*** / Questionnaire items | F1 | F2 | F3 | F4 | F5 | F6 | F7 | F8 |
|---|---|---|---|---|---|---|---|---|
| ***Skills (All sample: Cronbach's alpha= 0.783)*** | | | | | | | | |
| Q1 I often cook fresh food at home. | 0.376 | | | | | | | |
| Q2 I can prepare food quickly and efficiently | 0.793 | | | | | | | |
| Q3 I can cook using unfamiliar ingredients | 0.612 | | | | | | | |
| Q4 I think my cooking skills are good enough | 0.939 | | | | | | | |
| Q5 I enjoy spending time and effort on cooking and choosing ingredients. | 0.424 | | | | | | | |
| ***Environmentally-friendly Choices (All sample: Cronbach's alpha= 0.775)*** | | | | | | | | |
| Q6 I am conscious of environmentally and socially friendly actions | | -0.599 | | | | | | |
| Q7 I prefer to make environmentally friendly choices and actions. | | -0.556 | | | | | | |
| ***Attitude towards Environmental Activities (All sample: Cronbach's alpha= 0.764)*** | | | | | | | | |
| Q8 It is good to donate money or support organizations that contribute to the environment and society. | | | 0.861 | | | | | |
| Q9 It's good to be active in groups that contribute to the environment and society. | | | 0.741 | | | | | |
| ***Attitude towards Health (All sample: Cronbach's alpha= 0.825)*** | | | | | | | | |
| Q10 Living a healthy life improves performance. | | | | 0.573 | | | | |
| Q11 Living a healthy lifestyle is fun. | | | | 0.631 | | | | |
| Q12 It's good to lead a healthy lifestyle. | | | | 0.728 | | | | |
| Q13 I prefer to make healthy choices and take care of my health. | | | | 0.621 | | | | |
| ***Family/Peer Environment and Value (All sample: Cronbach's alpha= 0.723)*** | | | | | | | | |
| Q14 My close friends, family and acquaintances are understanding of the food and ingredients I choose. | | | | | -0.644 | | | |
| Q15 The health and preferences of my close friends, family and acquaintances influence my food choices and ingredients. | | | | | -0.804 | | | |
| Q16 I share my food preferences and values with my close friends, family and loved ones | | | | | -0.819 | | | |

| Factor / Questionnaire items | | | |
|---|---|---|---|
| ***Connection and Attachment to the Community (All sample: Cronbach's alpha= 0.806)*** | | | |
| Q17 I feel a sense of pride and attachment to the area where I live now. | -0.990 | | |
| Q18 The area where I live now is very important to me. | -0.825 | | |
| Q19 I like talking to people about the area where I live now. | -0.571 | | |
| Q20 I want the area where I live now to continue to exist in the future. | -0.608 | | |
| ***Culture (All sample: Cronbach's alpha= 0.777)*** | | | |
| Q21 The area where I live now has traditional scenery. | | -0.931 | |
| Q22 There is traditional culture in the area where I live now. | | -0.921 | |
| ***Local Norms (All sample: Cronbach's alpha= 0.782)*** | | | |
| Q23 The people in the area where I live think that supporting local farmers is a good thing. | | | 0.808 |
| Q24 The people in the area where I live would be happy if I bought local produce (such as produce from within the prefecture). | | | 0.728 |
| Q25 The people in the area where I live are trying to actively buy local produce (such as produce from within the prefecture). | | | 0.833 |
| Q26 Buying local produce (e.g. from within the prefecture) is a good thing. | | | 0.453 |

Note: KMO .935, Bartlett's test $p < 0.001$, Scale used: 1 (strongly disagree) to 7 (strongly agree). Principal axis factoring. Oblimin rotation (delta = 0.15) with Kaiser normalization. Absolute values of factor loading less than 0.30 are not shown.

Table 6 Pattern matrix of exploratory factor analysis for Seesaw of gain and loss

| ***Factor*** / Questionnaire items | F1 | F2 | F3 | F4 | F5 |
|---|---|---|---|---|---|
| ***Food Education and Learning Opportunities (All sample: Cronbach's alpha= 0.866)*** | | | | | |
| Q1 I want to have experiences and learn about food | 0.824 | | | | |
| Q2 I want to have experiences and learn about agriculture | 0.898 | | | | |
| Q3 I want to experience and learn about food production and plant cultivation | 0.881 | | | | |
| Q4 I want to interact with farmers in the area where I live | 0.617 | | | | |
| Q5 I want to have access to a wide variety of ingredients. | 0.367 | | | | |
| ***Contribution to Environmental and Social Issues (All sample: Cronbach's alpha= 0.850)*** | | | | | |
| Q6 I want to participate in activities that contribute to the environment and society | | 0.557 | | | |
| Q7 I want to support activities that contribute to the environment and society | | 0.9 | | | |
| Q8 I want to support environmentally and socially conscious food production and distribution. | | 0.865 | | | |
| Q9 I want to buy food produced in an environmentally and socially conscious way. | | 0.628 | | | |
| ***Connection with People (All sample: Cronbach's alpha= 0.719)*** | | | | | |
| Q10 I want to deepen my interactions with the people in my community. | | | 0.545 | | |
| Q11 I want to deepen my love for my family and close friends | | | 0.541 | | |
| Q12 I want to deepen my friendship with my friends. | | | 0.806 | | |
| ***Connection with Nature (All sample: Cronbach's alpha= 0.918)*** | | | | | |
| Q13 I want to spend more time in nature. | | | | -0.95 | |
| Q14 I want to increase my opportunities to be in touch with nature | | | | -0.852 | |
| Q15 I want to relax in nature. | | | | -0.957 | |
| Q16 I want to relieve stress by being in nature. | | | | -0.903 | |
| ***Variety of Ingredients (All sample: Cronbach's alpha= 0.815)*** | | | | | |
| Q17 I want to get my hands on fresh ingredients | | | | | 0.75 |
| Q18 I want to get my hands on delicious ingredients. | | | | | 0.711 |

Note: KMO .935, Bartlett's test $p < 0.001$, Scale used: 1 (strongly disagree) to 7 (strongly agree). Principal axis factoring. Oblimin rotation (delta = 0.1) with Kaiser normalization. Absolute values of factor loading less than 0.30 are not shown.

#### 4.2.2 Measurement model assessment

The PLS-SEM measurement model was evaluated using several indicators (J. Hair and Alamer 2022). First, outer loadings were calculated for all samples and segments. Observed variables constituting "Skills" showed inconsistent values across segments, with some meeting the 0.7 threshold and others not, indicating heterogeneity across segments. Consequently, Skills was excluded from the PLS-SEM model. Other observed variables were similarly assessed for outer loadings, and those with loadings below 0.7 were also excluded (Table A6).

Subsequently, the reliability and validity of the constructs were assessed. For "Connections with People," "Family/Peer Environment and Values," and "Local Norms," Cronbach's α and rho_a fell below 0.7 in some segments. However, as some studies suggest that reliability in the range of 0.6-0.7 is acceptable for exploratory research (Janssens et al. 2008; Joe F. Hair, Ringle, and Sarstedt 2011; Joseph F. Hair et al. 2019), these factors were supported. All VIF values were below the threshold of 5, indicating no multicollinearity concerns. These values are shown in Table A7 and A8. Next, using a two-step process, "Attitude towards Environmental Activities" and "Environmentally-friendly Choices" were constructed as "Attitude towards Environment" in the PLS-SEM model and evaluated using the same steps. In some segments, Cronbach's α and rho_a for "Attitude towards Environment" fell below 0.7, but were retained as above. From the results, H16 to H25 related to "Attitude towards Food" and Skills, and H28, H29, H32, H33, H37, H38, H42 and H43 were deleted, and the remaining hypotheses were validated. Finally, the discriminant validity was evaluated and confirmed the discriminant validity of the constructs (Table A9).

### 4.2.3 Path coefficients and findings

Figure 4 and Table 7 present the visualization of PLS-SEM for the entire sample group and summarize the path coefficients for all segments. The verified hypotheses H1, H6, H7, H9, H10, H13, H14, H15, H26, H27, H40, H41, and H44 were supported by statistically significant path coefficients ($p < 0.05$) across all segments, while other hypotheses were not significant in some or all segments.

Among the hypotheses directly related to CSA participation intention in the "Seesaw of gain and loss," H1, H2, and H4 were supported. H1 was significant across all segments, indicating that the educational benefits associated with CSA participation are an important factor. H2 was significant in all groups except "Farmers' Market Advocates," with "Sustainable Food Seekers" showing the highest value. This relationship suggests that many participants place importance on contributing to environmental and social issues

H4 was only significant for "Conventional Shoppers," but with a small value of -0.109 ($p < 0.05$), suggesting minimal influence. H6 and H7 were related to H1 and H2 across all segments. "Attitude towards Health" influenced other factors of the "Seesaw of gain and loss" (H3, H4, and H5), with H13 having path coefficients exceeding 0.3 across all segments. These findings suggest that "Attitude towards Environment" and "Attitude towards Health" are important factors related to the "Seesaw of gain and loss".

In addition, both are "Attitude towards Environment" and "Attitude towards Health" influenced by the "Family/Peer" and "Local Environment and Community" levels. Particularly for H40, H41, and H44, path

94 coefficients exceeded 0.3 in several segments, with “Sustainable Food
95 Seekers” showing the highest values among segments, suggesting that the
96 presence or absence of “Local Norms” indirectly influences CSA
97 participation. These findings highlight that motivations and perceived
98 benefits differ across segments, emphasizing the need for tailored
99 approaches to effectively engage each segment

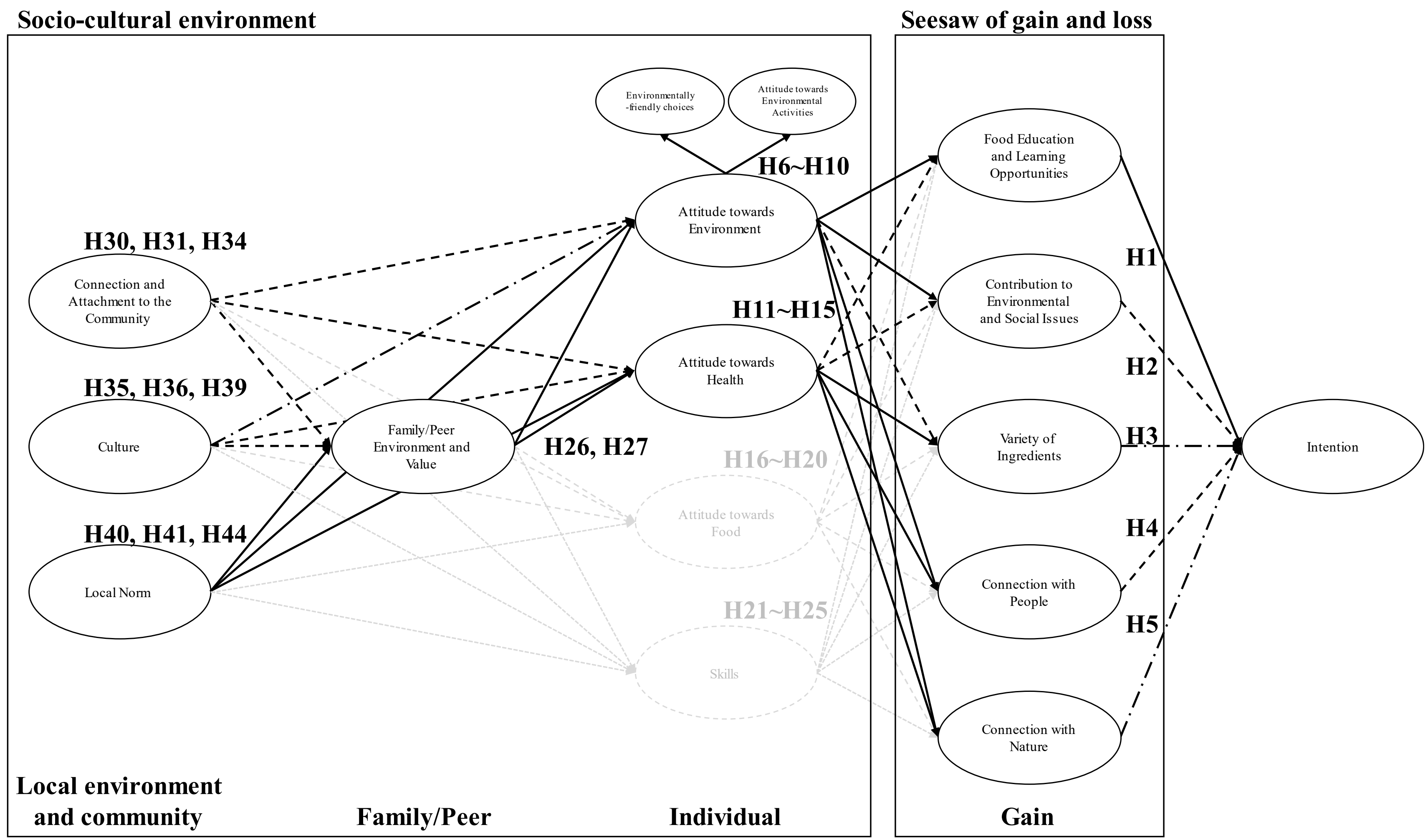

Fig. 4 Revised hypothesized model. The gray paths have not been verified in this study. The solid lines are significant ($p < 0.05$) for all sample groups and all segments. The dotted lines are significant ($p < 0.05$) for all sample groups or some segments. The long dash dotted lines are not significant for all segments.

104 Table 7 Path coefficient of the all-sample hypothesized model and each consumer segment.

| | All-Sample | Segment1 | Segment2 | Segment3 | Segment4 | Segment5 |
|---|---|---|---|---|---|---|
| H1 Food Education and Learning Opportunities → Intention | 0.398* | 0.42* | 0.372* | 0.439* | 0.367* | 0.329* |
| H2 Contribution to Environmental and Social Issues → Intention | 0.233* | 0.286* | 0.19* | 0.075 | 0.257* | 0.305* |
| H3 Variety of Ingredients → Intention | -0.001 | 0.018 | -0.031 | 0.104 | 0.083 | -0.042 |
| H4 Connection with People → Intention | 0.004 | -0.109* | 0.05 | -0.072 | -0.053 | 0.027 |
| H5 Connection with Nature → Intention | 0.049 | -0.019 | 0.051 | 0.074 | 0.109 | 0.049 |
| H6 Attitude towards Environment → Food Education and Learning Opportunities | 0.419* | 0.357* | 0.427* | 0.283* | 0.497* | 0.418* |
| H7 Attitude towards Environment → Contribution to Environmental and Social Issues | 0.658* | 0.656* | 0.703* | 0.625* | 0.593* | 0.614* |
| H8 Attitude towards Environment → Variety of Ingredients | 0.075* | 0.058 | 0.011 | 0.094 | 0.168* | 0.118* |
| H9 Attitude towards Environment → Connection with People | 0.333* | 0.297* | 0.382* | 0.152* | 0.357* | 0.302* |
| H10 Attitude towards Environment → Connection with Nature | 0.27* | 0.281* | 0.243* | 0.155* | 0.352* | 0.257* |
| H11 Attitude towards Health → Food Education and Learning Opportunities | 0.035 | 0.107 | -0.005 | 0.214* | -0.079 | 0.044 |
| H12 Attitude towards Health → Contribution to Environmental and Social Issues | 0.063* | 0.023 | 0.051 | 0.056 | 0.131 | 0.098* |
| H13 Attitude towards Health → Variety of Ingredients | 0.594* | 0.525* | 0.598* | 0.497* | 0.578* | 0.607* |
| H14 Attitude towards Health → Connection with People | 0.315* | 0.197* | 0.272* | 0.435* | 0.309* | 0.407* |
| H15 Attitude towards Health → Connection with Nature | 0.307* | 0.209* | 0.302* | 0.39* | 0.187* | 0.4* |
| H26 Family/Peer Environment and Value → Attitude towards Environment | 0.206* | 0.197* | 0.18* | 0.209* | 0.155* | 0.254* |
| H27 Family/Peer Environment and Value → Attitude towards Health | 0.205* | 0.264* | 0.158* | 0.269* | 0.246* | 0.175* |
| H30 Connection and Attachment to the Community → Attitude towards Environment | 0.188* | 0.126 | 0.222* | 0.226* | 0.21* | 0.155* |
| H31 Connection and Attachment to the Community → Attitude towards Health | 0.212* | 0.11 | 0.285* | 0.109 | 0.392* | 0.144* |
| H34 Connection and Attachment to the Community → Family/Peer Environment and Value | 0.182* | 0.166* | 0.191* | 0.172* | 0.074 | 0.245* |
| H35 Culture → Attitude towards Environment | 0.011 | -0.005 | 0.039 | -0.054 | -0.016 | 0.041 |
| H36 Culture → Attitude towards Health | -0.074* | -0.035 | -0.117* | -0.096 | -0.015 | -0.027 |
| H39 Culture → Family/Peer Environment and Value | 0.109* | 0.003 | 0.086 | 0.196* | 0.242* | 0.107* |
| H40 Local Norm → Attitude towards Environment | 0.418* | 0.391* | 0.358* | 0.424* | 0.441* | 0.456* |
| H41 Local Norm → Attitude towards Health | 0.417* | 0.284* | 0.353* | 0.46* | 0.26* | 0.564* |
| H44 Local Norm → Family/Peer Environment and Value | 0.335* | 0.294* | 0.326* | 0.211* | 0.32* | 0.366* |

105 (Segment1: Conventional Shoppers, Segment2: Organic Enthusiasts, Segment3: Farmers' Market Advocates, Segment4: Home Delivery
106 Preferers, Segment5: Sustainable Food Seekers)
107 Significance codes: * < 0.05 (two-tailed). Percentile bootstrap with 10000 resampling.

### 4.3 Awareness and experience of CSA in Japan

The results of CSA awareness are presented in Table 3. 75.4% of respondents were unaware of both the term and explanation of CSA. Only 3.5% of respondents knew both the term and its explanation. Among those who were aware of CSA, 2.8% had participated in CSA. In comparison, a nationwide survey in the United States found that approximately 21.6% of respondents were past or current CSA members (J. Chen et al. 2019). These findings highlight that CSA is rarely practiced in Japan, and there are limited opportunities for participation.

Table 3 Response regarding CSA awareness

| ***Question***/Response | *N* | % |
|---|---|---|
| ***Were you aware of CSA before answering this questionnaire?*** | | |
| I didn't know the word and the explanation before answering this questionnaire. | 1873 | 75.4% |
| I only knew the word before answering the questionnaire, but I didn't know the explanation. | 371 | 14.9% |
| I only knew the explanation before answering the questionnaire, but I didn't know the word. | 154 | 6.2% |
| I knew the word and the explanation before answering the questionnaire. | 86 | 3.5% |
| ***Please tell us about your experience with CSA. (for everyone except those who answered "I didn't know the word and the explanation before answering this questionnaire." above).*** | | |
| I have never participated and have never considered participating. | 476 | 77.9% |
| I have never participated, but I have considered participating. | 118 | 19.3% |
| I have participated. | 17 | 2.8% |

## 5. Discussion

### 5.1 Potential CSA participants in Japan and their characteristics

In the LCA, "Sustainable Food Seekers" was the only segment showing positive utility for CSA, suggesting that consumers in this segment are potential CSA participants in Japan. Previous research identified young generation, female, and high-income consumers as typical CSA participants. Our results align with these findings in terms of younger age and higher income compared to other segments (J. Chen et al. 2019; Bernard, Bonein, and Bougherara 2020). However, the gender distribution within this segment was nearly equal, differing from previous studies (Table A10 and A11).

While many previous studies cited access to organically grown vegetables as an important factor for CSA (Perez, Allen, and Brown 2003; Hvitsand 2016; Galt et al. 2019; Bernard, Bonein, and Bougherara 2020), "Sustainable Food Seekers" showed no utility for organic certification attributes. Furthermore, the largest segment, "Organic Enthusiasts," showed no utility for CSA. Regarding CSA awareness, 40.7% of "Sustainable Food Seekers" and 23.3% of "Organic Enthusiasts" were familiar with both the term and explanation. Among respondents who had considered CSA participation, 53% were Sustainable Food Seekers," while only 7.6% were "Organic Enthusiasts." This suggests that while "Organic Enthusiasts" are aware of CSA, they find little appeal beyond access to organically grown ingredients.

In recent years, organic vegetables and foods have become readily available in supermarkets and specialty stores in Japan, allowing consumers to choose organic products daily. This accessibility may mean that the time and financial losses associated with CSA participation do not balance out the benefit of obtaining organic vegetables. Previous research has also shown that while some Japanese consumers have a special interest in organic vegetables (Ujiie and Matsuoka 2021), many consumers prioritize other food attributes such as safety certification and freshness (Yang et al. 2021; Sajiki and Lu 2022). This suggest that not only are organic ingredients widely available in the Japanese market, but consumers also prioritize factors such as origin and freshness, potentially reducing the significance of CSA as a source of organic vegetables in Japan compared to other countries.

These findings emphasize the need to clearly demonstrate the value of CSA in "Contribution to Environmental and Social Issues" rather than focusing on access to "Variety of Ingredients", as mentioned in previous research, to promote CSA activities in Japan targeting potential participants. This approach is particularly important given the niche status of CSA in Japan.

### 5.2 Reconsidering the CSA model in Japan

This study's results indicate that CSA participation intention was influenced by H1 "Food Education and Learning Opportunities" across all segments and H2 "Contribution to Environmental and Social Issues" in some segments. This differs from previous research which identified H3 "Variety of Ingredients" as the main factor (Takagi et al. 2024). This difference may be attributed to Japan's unique approach to food education. Japan is the only country with a national law on food education (Basic Act on Shokuiku (Food and Nutrition Education), enacted in 2005). This law promotes food education activities nationwide, emphasizing physical and mental health through food, understanding of traditional food culture, and harmony between food, environment, and society. While the USDA in America and CHAFEA in the EU also promote food education through food guides, Japan's national-level approach is distinctive, potentially explaining why H1 and H2 emerged as crucial factors for CSA participation intention.

Japanese consumers' high concern for product safety likely influenced these results. Previous research comparing Japan, Taiwan, and Indonesia showed that Japanese consumers prioritize food safety the most (Yang et al. 2021). In addition, it is also possible that H3 was not the main factor due to the interest of Japanese consumers in food and the environment for purchasing ingredients. JETRO and USDA noted increased attention to food safety among Japanese consumers following pesticide residue issues in imported frozen foods from 2008 to 2009 (JETRO 2011; USDA 2016). A survey of Japanese university students revealed low evaluations of imported food quality and safety (Kato 2010), and in a survey of overseas consumers conducted by JETRO in seven countries, 10.7% prioritized "safety" when purchasing Japanese food products (JETRO 2013). These findings indicate a very high awareness of food safety in Japan and a global recognition of the safety of Japanese food ingredients.

The fact that Japan has an environment where consumers can easily obtain a wide variety of safe and domestic ingredients on a daily basis may be the reason why "Variety of Ingredients," including Q17 "I want to get my hands on fresh ingredients" and Q18 "I want to get my hands on delicious ingredients.", did not have a significant impact on the intention to participate in CSA.

Furthermore, the study's finding that "Attitude towards Health" (H13) most strongly influenced the "Variety of Ingredients." These results and the Japanese consumers' awareness of safety suggest that the "Attitude towards Health" includes not only a healthy diet and lifestyle, but also a desire to avoid harmful additives and pollutants that can lead to health problems.

Across all segments, "Attitude towards Environment" (H6 and H7) influenced the two gains: H1 "Food Education and Learning Opportunities" and H2 "Contribution to Environmental and Social issues." This suggests that fostering individual-level "Attitude towards Environment" plays a critical role in promoting CSA participation among Japanese consumers. However, previous research indicates that Japanese people already have a high level of environmental attitude (Jussaume and Higgins 1998). Therefore, to promote CSA, it may be important to present CSA participation as a concrete action for solving environmental and social issues.

In all segments, "Attitude towards Environment" and "Attitude towards Health" were influenced by H27 "Family/Peer Environment and Value" and H41 "Local Norms". For "Attitude towards Environment," this similar with Turkish research showing that strengthening mother-daughter communication increases intergenerational similarity in sustainable consumption attitudes and behaviors (Essiz and Mandrik 2022), and Japanese research demonstrating that social norm evaluation is a key factor in environmentally conscious behaviors visible to neighbors, such as recycling and watering garden trees (Hirose 1994). Regarding "Attitude towards Health", this is consistent with Japanese research revealing a correlation between mothers providing healthy meals at home and health literacy (Yoshii et al. 2021), and studies suggesting that social capital in small Japanese communities plays a crucial role in promoting health among the elderly (Ichida et al. 2009). These insights suggest that promoting CSA in Japan might be most effective by emphasizing its educational aspects and approaching families as units, or by implementing CSA as a visible and easily understandable environmentally and socially conscious practice within local communities.

Previous research suggested a reflexive process where participation updates the socio-cultural environment and the balance of benefits and losses, influencing individual decision-making (Takagi et al. 2024). While the papers reviewed in previous research focused on factors for participation among existing CSA participants, the majority of respondents in this study have not participated in CSA. Given these differences, future research should investigate how expected gains change before and after participation.

### 5.3 A multidimensional approach to promoting CSA

This study's findings offer important implications for various stakeholders interested in implementing and promoting CSA activities.

For policymakers, there is a need to recognize the environmental and social value of CSA and develop systems to support its dissemination. Given the close alignment between the principles of Japan's "Basic Act on Shokuiku" and CSA activities, it could be effective to promote CSA as part of food education initiatives. Previous research has shown that the existence of support systems and organizations for CSA can be a factor in its implementation (Mert-Cakal and Miele 2020). Moreover, the production costs of organic farming, which is practiced by many CSA farmers, are about 50-100% higher than conventional farming (Håkansson, Ascard, and Söderlind 2009), making it difficult to achieve cost benefits from CSA practice despite the prepayment system. Therefore, policymakers could consider subsidy programs for CSA organizations and farmers, while commissioning experiential learning programs at CSA farms as part of food education promotion measures. This could contribute to achieving various objectives of the Basic Act on Food Education, including promoting understanding of food and agriculture through CSA, encouraging local production for local consumption, and linking with environmental education.

For CSA implementing organizations, developing marketing strategies targeting the "Sustainable Food Seekers" segment identified in this study could be effective. For this segment, emphasizing the value of CSA in "Contributing to Environmental and Social Issues" is likely to be impactful. As previous research suggests, building cooperative relationships with environmental and social contribution organizations could potentially strengthen CSA initiatives (Bonfert 2022). Furthermore, respecting local values and norms and organizing events and workshops involving the entire community could be effective in promoting CSA.

For producers, it is necessary to clearly communicate the environmental and social value of CSA and emphasize the importance of building direct relationships with consumers. As this study revealed that consumer's intentions to participate in CSA are influenced by "Food Education and Learning Opportunities," sharing production processes and agricultural challenges with consumers through social media and farm tours could help bridge the gap with consumer perceptions.

Implementing these recommendations based on the insights gained from this study is expected to significantly contribute to the spread of CSA in Japan and the construction of sustainable food systems. Simultaneously, the study's finding of remarkably low awareness of CSA strongly suggests the need for continued efforts to enhance consumer understanding.

In the future, cooperation among various stakeholders to spread the value of CSA throughout society will be an important step toward building sustainable agriculture and food systems.

## 6. Conclusion

This study aimed to identify potential CSA participants in Japan and elucidate the factors influencing Japanese consumers' willingness to participate in CSA through quantitative verification of CSA participation model. To achieve this, we conducted LCA based on a choice experiment and quantitative verification of the CSA participation model using PLS-SEM. The choice experiment results revealed that the "Sustainable Food Seekers" segment represents the most promising group for CSA participation in Japan. This segment exhibited the youngest average age, with statistically significant differences observed between this group and certain other segments. Furthermore, significant differences were also found in terms of annual individual income levels when compared to other segments.

Additionally, the quantitative validation of the CSA participation model highlighted that intentions to participate in CSA are strongly associated with "Food Education and Learning Opportunities" and "Contribution to Environmental and Social Issues." Notably, "Sustainable Food Seekers" placed greater emphasis on the "Contribution to Environmental and Social Issues" compared to other segments. These findings contrast with factors identified in other countries, such as access to organic or diverse produce. This distinction can be attributed to Japan's unique food culture, food education, high food safety standards, and the wide availability of diverse, high-quality agricultural products in conventional markets.

From a socio-cultural perspective, the results suggest that fostering individual-level "Attitude towards Environment" plays a crucial role in promoting CSA participation. These attitudes are shaped by the influence of "Family/Peer Environment and Values" as well as "Local Norms." These insights indicate that the most effective strategies for promoting CSA in Japan involve emphasizing its educational aspects and approaching families as units. Alternatively, CSA could be promoted as a visible and easily understandable initiative for environmental and social contribution within local communities.

However, the low overall awareness of CSA in Japan presents a significant challenge. Future efforts should focus on increasing public awareness about CSA and its potential contributions to sustainable agriculture and community development. It may also be important for policy makers to integrate CSA into existing food education initiatives in line with Japan's Basic Act on Shokuiku.

Our research also underscores the importance of considering cultural and market contexts when promoting alternative food networks like CSA. The strategies that have proven successful in other countries may need to be adapted to suit the unique characteristics of the Japanese market and consumer preferences.

While this study provides valuable insights, it also has limitations. First, our research focused on Japanese consumers, which may limit the generalizability of our findings to other cultural contexts. Second, the study primarily relied on self-reported data, which may be subject to social desirability bias. Future research could benefit from incorporating observational or interview data to complement survey responses. Additionally, our study design limits our ability to capture changes in consumer attitudes and behaviors over time. Longitudinal studies could provide a more dynamic understanding of CSA participation patterns.

In conclusion, while CSA faces challenges in gaining widespread adoption in Japan, there is potential for growth if properly aligned with consumer values and societal needs. Future research could explore the long-term impacts of CSA participation on consumer behavior and local food systems in Japan, as well as investigate effective strategies for scaling up CSA initiatives while maintaining their core principles of community engagement and sustainable agriculture.

## 7. Declaration of AI and AI-assisted technologies in the writing process

In writing this paper, after preparing a full text draft in the lead author's non-English native language, the authors used OpenAI's artificial intelligence language model, ChatGPT, to prepare an English draft. After using this tool, the authors carefully reviewed and edited the generated content to ensure the flow, logic, and accuracy of the text, making additions as necessary. Therefore, full responsibility for the content of the publication rests with the authors.

8. CRediT authorship contribution statement

Sota Takagi: Conceptualization, Methodology, Resources, Formal analysis, Investigation, Writing – Original Draft, Visualization. Miki Saijo: Validation, Writing - Review & Editing. Takumi Ohashi: Supervision, Project administration, Validation, Writing - Review & Editing, Funding acquisition. All authors contributed critically to the drafts and gave final approval for publication.

9. Acknowledgments

This work was supported by MAFF Commissioned project study on “Development of comfortable management systems for poultry and swine: Grant Number JPJ011279.”

10. Conflict of interest

The authors declare no conflicts of interest associated with this manuscript.

## Appendix

Table A1. Questions and References (Socio cultural environment)

| ***Factor*** / Questionnaire items | | References |
|---|---|---|
| ***Skills*** | | |
| | Q1 I often cook fresh food at home. | (Vasquez et al. 2016) |
| | Q2 I can prepare food quickly and efficiently | |
| | Q3 I can cook using unfamiliar ingredients | (Vasquez et al. 2016) |
| | Q4 I think my cooking skills are good enough | (Hartmann, Dohle, and Siegrist 2013) |
| | Q5 I enjoy spending time and effort on cooking and choosing ingredients. | (Wenzig and Gruchmann 2018) |
| ***Environmentally-friendly Choices*** | | |
| | Q6 I am conscious of environmentally and socially friendly actions | (Khare 2015) |
| | Q7 I prefer to make environmentally friendly choices and actions. | (Khare 2015; Vasquez et al. 2016) |
| ***Attitude towards Environmental Activities*** | | |
| | Q8 It is good to donate money or support organizations that contribute to the environment and society. | (Khare 2015) |
| | Q9 It's good to be active in groups that contribute to the environment and society. | (Khare 2015) |
| ***Attitude towards Health*** | | |
| | Q10 Living a healthy life improves performance. | (Roininen, Lähteenmäki, and Tuorila 1999; Salmani et al. 2020) |
| | Q11 Living a healthy lifestyle is fun. | (Roininen, Lähteenmäki, and Tuorila 1999; Salmani et al. 2020) |
| | Q12 It's good to lead a healthy lifestyle. | (Roininen, Lähteenmäki, and Tuorila 1999; Salmani et al. 2020) |
| | Q13 I prefer to make healthy choices and take care of my health. | (Roininen, Lähteenmäki, and Tuorila 1999; Salmani et al. 2020) |
| ***Family/Peer Environment and Value*** | | |
| | Q14 My close friends, family and acquaintances are understanding of the food and ingredients I choose. | (Crandall et al. 2020) |
| | Q15 The health and preferences of my close friends, family and acquaintances influence my food choices and ingredients. | (Crandall et al. 2020) |
| | Q16 I share my food preferences and values with my close friends, family and loved ones | (Crandall et al. 2020) |
| ***Connection and Attachment to the Community*** | | |
| | Q17 I feel a sense of pride and attachment to the area where I live now. | (Kyle, Graefe, and Manning 2005) |
| | Q18 The area where I live now is very important to me. | (Kyle, Graefe, and Manning 2005) |
| | Q19 I like talking to people about the area where I live now. | (Scopelliti and Tiberio 2010) |
| | Q20 I want the area where I live now to continue to exist in the future. | (Scopelliti and Tiberio 2010) |
| ***Culture*** | | |
| | Q21 The area where I live now has traditional scenery. | (Bernardo, Loupa-Ramos, and Coelho 2023) |
| | Q22 There is traditional culture in the area where I live now. | (Bernardo, Loupa-Ramos, and Coelho 2023) |
| ***Local Norms*** | | |
| | Q23 The people in the area where I live think that supporting local farmers is a good thing. | (Al-Swidi et al. 2014) |
| | Q24 The people in the area where I live would be happy if I bought local produce (such as produce from within the prefecture). | (Al-Swidi et al. 2014) |
| | Q25 The people in the area where I live are trying to actively buy local produce (such as produce from within the prefecture). | (Al-Swidi et al. 2014) |
| | Q26 Buying local produce (e.g. from within the prefecture) is a good thing. | (Al-Swidi et al. 2014) |

Table A2. Questions and References (Seesaw of gain and loss)

| ***Factor*** / Questionnaire items | | References |
|---|---|---|
| ***Food Education and Learning Opportunities*** | | |
| | Q1 I want to have experiences and learn about food | (Vasquez et al. 2016) |
| | Q2 I want to have experiences and learn about agriculture | (W. Chen 2013) |
| | Q3 I want to experience and learn about food production and plant cultivation | (W. Chen 2013) |
| | Q4 I want to interact with farmers in the area where I live | (Gugerell et al. 2021) |
| | Q5 I want to have access to a wide variety of ingredients. | (J. Chen et al. 2019) |
| ***Contribution to Environmental and Social Issues*** | | |
| | Q6 I want to participate in activities that contribute to the environment and society | (Milfont and Duckitt 2010) |
| | Q7 I want to support activities that contribute to the environment and society | (Milfont and Duckitt 2010) |
| | Q8 I want to support environmentally and socially conscious food production and distribution. | (Milfont and Duckitt 2010) |
| | Q9 I want to buy food produced in an environmentally and socially conscious way. | (Tandon et al. 2020) |
| ***Connection with People*** | | |
| | Q10 I want to deepen my interactions with the people in my community. | (Gugerell et al. 2021) |
| | Q11 I want to deepen my love for my family and close friends | (W. Chen 2013) |
| | Q12 I want to deepen my friendship with my friends. | (W. Chen 2013) |
| ***Connection with Nature*** | | |
| | Q13 I want to spend more time in nature. | (Milfont and Duckitt 2010) |
| | Q14 I want to increase my opportunities to be in touch with nature | (Milfont and Duckitt 2010) |
| | Q15 I want to relax in nature. | (Milfont and Duckitt 2010) |
| | Q16 I want to relieve stress by being in nature. | (Milfont and Duckitt 2010) |
| ***Variety of Ingredients*** | | |
| | Q17 I want to get my hands on fresh ingredients | (W. Chen 2013) |
| | Q18 I want to get my hands on delicious ingredients. | (W. Chen 2013) |

604

605 Table A3. Hypotheses of proposed conceptual model

| Hypotheses | Description |
|---|---|
| H1 Food Education and Learning Opportunities → Intention | Food Education and Learning Opportunities positively influence the intention to participate. |
| H2 Contribution to Environmental and Social Issues → Intention | Contribution to Environmental and Social Issues positively influences the intention to participate. |
| H3 Variety of Ingredients → Intention | Variety of Ingredients positively influences the intention to participate. |
| H4 Connection with People → Intention | Connection with People positively influences the intention to participate. |
| H5 Connection with Nature → Intention | Connection with Nature positively influences the intention to participate. |
| H6 Attitude towards Environment → Food Education and Learning Opportunities | Attitude towards the Environment positively influences food education and learning opportunities. |
| H7 Attitude towards Environment → Contribution to Environmental and Social Issues | Attitude towards the Environment positively influences the contribution to environmental and social issues. |
| H8 Attitude towards Environment → Variety of Ingredients | Attitude towards the Environment positively influences the variety of ingredients. |
| H9 Attitude towards Environment → Connection with People | Attitude towards the Environment positively influences the connection with people. |
| H10 Attitude towards Environment → Connection with Nature | Attitude towards the Environment positively influences the connection with nature. |
| H11 Attitude towards Health → Food Education and Learning Opportunities | Attitude towards Health positively influences food education and learning opportunities. |
| H12 Attitude towards Health → Contribution to Environmental and Social Issues | Attitude towards Health positively influences the contribution to environmental and social issues. |
| H13 Attitude towards Health → Variety of Ingredients | Attitude towards Health positively influences the variety of ingredients. |
| H14 Attitude towards Health → Connection with People | Attitude towards Health positively influences the connection with people. |
| H15 Attitude towards Health → Connection with Nature | Attitude towards Health positively influences the connection with nature. |
| H16 Attitude towards Food → Food Education and Learning Opportunities | Attitude towards Food positively influences food education and learning opportunities. |
| H17 Attitude towards Food → Contribution to Environmental and Social Issues | Attitude towards Food positively influences the contribution to environmental and social issues. |
| H18 Attitude towards Food → Variety of Ingredients | Attitude towards Food positively influences the variety of ingredients. |
| H19 Attitude towards Food → Connection with People | Attitude towards Food positively influences the connection with people. |
| H20 Attitude towards Food → Connection with Nature | Attitude towards Food positively influences the connection with nature. |
| H21 Skills → Food Education and Learning Opportunities | Skills positively influence food education and learning opportunities. |
| H22 Skills → Contribution to Environmental and Social Issues | Skills positively influence the contribution to environmental and social issues. |
| H23 Skills → Variety of Ingredients | Skills positively influence the variety of ingredients. |
| H24 Skills → Connection with People | Skills positively influence the connection with people. |
| H25Skills → Connection with Nature | Skills positively influence the connection with nature. |
| H26 Family/Peer Environment and Value → Attitude towards Environment | Family/Peer Environment and Values positively influence the attitude towards the environment. |
| H27 Family/Peer Environment and Value → Attitude towards Health | Family/Peer Environment and Values positively influence the attitude towards health. |
| H28 Family/Peer Environment and Value → Attitude towards Food | Family/Peer Environment and Values positively influence the attitude towards food. |
| H29 Family/Peer Environment and Value → Skills | Family/Peer Environment and Values positively influence skills. |
| H30 Connection and Attachment to the Community → Attitude towards Environment | Connection and Attachment to the Community positively influence the attitude towards the environment. |
| H31 Connection and Attachment to the Community → Attitude towards Health | Connection and Attachment to the Community positively influence the attitude towards health. |
| H32 Connection and Attachment to the Community → Attitude towards Food | Connection and Attachment to the Community positively influence the attitude towards food. |
| H33 Connection and Attachment to the Community → Skills | Connection and Attachment to the Community positively influence skills. |
| H34 Connection and Attachment to the Community → Family/Peer Environment and Value | Connection and Attachment to the Community positively influence family/peer environment and values. |
| H35 Culture → Attitude towards Environment | Culture positively influences the attitude towards the environment. |
| H36 Culture → Attitude towards Health | Culture positively influences the attitude towards health. |
| H37 Culture → Attitude towards Food | Culture positively influences the attitude towards food. |
| H38 Culture → Skills | Culture positively influences skills. |
| H39 Culture → Family/Peer Environment and Value | Culture positively influences family/peer environment and values. |
| H40 Local norm → Attitude towards Environment | Local norms positively influence the attitude towards the environment. |
| H41 Local norm → Attitude towards Health | Local norms positively influence the attitude towards health. |
| H42 Local norm → Attitude towards Food | Local norms positively influence the attitude towards food. |
| H43 Local norm → Skills | Local norms positively influence skills. |
| H44 Local norm → Family/Peer Environment and Value | Local norms positively influence family/peer environment and values. |

606

607

608 Table A4: AIC and BIC

| # Class | AIC | BIC |
|---|---|---|
| 3 | 45342.2 | 45912.33 |
| 4 | 44140.92 | 44903.03 |
| 5 | 42544.13 | 43498.22 |
| 6 | 42564.69 | 43710.76 |
| 7 | 41462.38 | 42800.44 |

Table A5: Actual and random data eigenvalues

| Factor | Actual data eigenvalues | Random data eigenvalues (mean) |
|---|---|---|
| ***Socio cultural environment*** | | |
| 1 | 8.268 | 0.201 |
| 2 | 2.002 | 0.174 |
| 3 | 1.240 | 0.152 |
| 4 | 0.573 | 0.135 |
| 5 | 0.529 | 0.118 |
| 6 | 0.404 | 0.103 |
| 7 | 0.315 | 0.089 |
| 8 | 0.222 | 0.075 |
| 9 | 0.133 | 0.062 |
| 10 | -0.008 | 0.049 |
| ***Seesaw of gain and loss*** | | |
| 1 | 7.866 | 0.157 |
| 2 | 1.545 | 0.130 |
| 3 | 1.011 | 0.108 |
| 4 | 0.395 | 0.089 |
| 5 | 0.323 | 0.072 |
| 6 | 0.039 | 0.056 |

Table A6: Outer loadings of factors used in PLS-SEM

| | All-Sample | Segment1 | Segment2 | Segment3 | Segment4 | Segment5 |
|---|---|---|---|---|---|---|
| ***Socio cultural environment*** | | | | | | |
| Attitude towards Environmental Activities ← Attitude towards Environment | 0.876 | 0.829 | 0.871 | 0.845 | 0.888 | 0.897 |
| Environmentally-friendly Choices ← Attitude towards Environment | 0.866 | 0.868 | 0.864 | 0.883 | 0.88 | 0.864 |
| Q6 ← Environmentally-Friendly Choices | 0.897 | 0.909 | 0.91 | 0.889 | 0.91 | 0.87 |
| Q7 ← Environmentally-Friendly Choices | 0.91 | 0.922 | 0.913 | 0.918 | 0.908 | 0.893 |
| Q8 ← Attitude towards Environmental Activities | 0.899 | 0.879 | 0.896 | 0.895 | 0.899 | 0.909 |
| Q9 ← Attitude towards Environmental Activities | 0.9 | 0.911 | 0.909 | 0.869 | 0.85 | 0.905 |
| Q10 ← Attitude towards Health | 0.823 | 0.805 | 0.802 | 0.803 | 0.773 | 0.855 |
| Q11 ← Attitude towards Health | 0.818 | 0.797 | 0.805 | 0.851 | 0.776 | 0.838 |
| Q12 ← Attitude towards Health | 0.821 | 0.776 | 0.808 | 0.794 | 0.812 | 0.854 |
| Q13 ← Attitude towards Health | 0.778 | 0.811 | 0.759 | 0.786 | 0.791 | 0.774 |
| Q14 ← Family/Peer Environment and Value | 0.832 | 0.825 | 0.842 | 0.835 | 0.795 | 0.836 |
| Q15 ← Family/Peer Environment and Value | 0.793 | 0.747 | 0.788 | 0.813 | 0.741 | 0.814 |
| Q16 ← Family/Peer Environment and Value | 0.783 | 0.799 | 0.781 | 0.728 | 0.73 | 0.808 |
| Q17 ← Connection and Attachment to the Community | 0.849 | 0.829 | 0.867 | 0.847 | 0.896 | 0.821 |
| Q18 ← Connection and Attachment to the Community | 0.817 | 0.851 | 0.828 | 0.815 | 0.775 | 0.812 |
| Q20 ← Connection and Attachment to the Community | 0.848 | 0.835 | 0.833 | 0.842 | 0.891 | 0.85 |
| Q21 ← Culture | 0.896 | 0.917 | 0.909 | 0.875 | 0.892 | 0.885 |
| Q22 ← Culture | 0.912 | 0.888 | 0.902 | 0.896 | 0.935 | 0.923 |
| Q23 ← Local Norm | 0.796 | 0.722 | 0.799 | 0.775 | 0.765 | 0.82 |
| Q24 ← Local Norm | 0.797 | 0.764 | 0.8 | 0.753 | 0.776 | 0.816 |
| Q26 ← Local Norm | 0.832 | 0.832 | 0.814 | 0.837 | 0.834 | 0.847 |
| ***Seesaw of gain and loss*** | | | | | | |
| Q1 ← Food Education and Learning Opportunities | 0.86 | 0.85 | 0.862 | 0.891 | 0.841 | 0.83 |
| Q2 ← Food Education and Learning Opportunities | 0.878 | 0.889 | 0.865 | 0.868 | 0.862 | 0.867 |
| Q3 ← Food Education and Learning Opportunities | 0.886 | 0.895 | 0.89 | 0.884 | 0.869 | 0.867 |
| Q4 ← Food Education and Learning Opportunities | 0.832 | 0.821 | 0.843 | 0.824 | 0.758 | 0.828 |
| Q6 ← Contribution to Environmental and Social Issues | 0.799 | 0.817 | 0.809 | 0.815 | 0.745 | 0.767 |
| Q7 ← Contribution to Environmental and Social Issues | 0.865 | 0.854 | 0.877 | 0.85 | 0.867 | 0.853 |
| Q8 ← Contribution to Environmental and Social Issues | 0.847 | 0.832 | 0.851 | 0.859 | 0.829 | 0.835 |
| Q9 ← Contribution to Environmental and Social Issues | 0.81 | 0.773 | 0.812 | 0.768 | 0.803 | 0.82 |
| Q10 ← Connection with People | 0.761 | 0.782 | 0.786 | 0.74 | 0.74 | 0.727 |
| Q11 ← Connection with People | 0.81 | 0.774 | 0.799 | 0.798 | 0.78 | 0.832 |
| Q12 ← Connection with People | 0.828 | 0.782 | 0.841 | 0.844 | 0.77 | 0.829 |
| Q13 ← Connection with Nature | 0.905 | 0.926 | 0.904 | 0.891 | 0.891 | 0.887 |
| Q14 ← Connection with Nature | 0.893 | 0.915 | 0.891 | 0.903 | 0.853 | 0.884 |
| Q15 ← Connection with Nature | 0.897 | 0.918 | 0.902 | 0.897 | 0.864 | 0.876 |
| Q16 ← Connection with Nature | 0.889 | 0.909 | 0.904 | 0.891 | 0.841 | 0.859 |
| Q17 ← Variety of Ingredients | 0.921 | 0.923 | 0.931 | 0.903 | 0.914 | 0.91 |
| Q18 ← Variety of Ingredients | 0.917 | 0.908 | 0.924 | 0.892 | 0.906 | 0.918 |

(Segment1: Conventional Shoppers, Segment2: Organic Enthusiasts, Segment3: Farmers' Market Advocates, Segment4: Home Delivery Preferers, Segment5: Sustainable Food Seekers)

Table A7: Construct reliability and validity

| ***Class***/Construct | Cronbach's alpha | Composite reliability (rho_a) | Composite reliability (rho_c) | Average variance extracted (AVE) |
|---|---|---|---|---|
| ***All-Sample*** | | | | |
| Food Education and Learning Opportunities | 0.887 | 0.888 | 0.922 | 0.747 |
| Contribution to Environmental and Social Issues | 0.85 | 0.852 | 0.899 | 0.69 |
| Variety of Ingredients | 0.815 | 0.816 | 0.916 | 0.844 |
| Connection with People | 0.718 | 0.72 | 0.842 | 0.64 |
| Connection with Nature | 0.918 | 0.918 | 0.942 | 0.803 |
| Attitude towards Environment | 0.683 | 0.683 | 0.863 | 0.759 |
| Attitude towards Health | 0.825 | 0.826 | 0.884 | 0.656 |
| Family/Peer Environment and Value | 0.725 | 0.734 | 0.844 | 0.644 |
| Connection and Attachment to the Community | 0.79 | 0.803 | 0.876 | 0.703 |
| Culture | 0.778 | 0.781 | 0.9 | 0.818 |
| Local norm | 0.737 | 0.745 | 0.85 | 0.654 |
| ***Conventional Shoppers*** | | | | |
| Food Education and Learning Opportunities | 0.887 | 0.888 | 0.922 | 0.747 |
| Contribution to Environmental and Social Issues | 0.837 | 0.838 | 0.891 | 0.672 |
| Variety of Ingredients | 0.808 | 0.812 | 0.912 | 0.839 |
| Connection with People | 0.681 | 0.685 | 0.823 | 0.607 |
| Connection with Nature | 0.937 | 0.937 | 0.955 | 0.84 |
| Attitude towards Environment | 0.612 | 0.617 | 0.837 | 0.72 |
| Attitude towards Health | 0.81 | 0.814 | 0.875 | 0.636 |
| Family/Peer Environment and Value | 0.701 | 0.711 | 0.833 | 0.625 |
| Connection and Attachment to the Community | 0.792 | 0.806 | 0.877 | 0.704 |
| Culture | 0.774 | 0.785 | 0.898 | 0.815 |
| Local norm | 0.677 | 0.715 | 0.817 | 0.599 |
| ***Organic Enthusiasts*** | | | | |
| Food Education and Learning Opportunities | 0.888 | 0.889 | 0.922 | 0.748 |
| Contribution to Environmental and Social Issues | 0.858 | 0.861 | 0.904 | 0.702 |
| Variety of Ingredients | 0.837 | 0.839 | 0.925 | 0.86 |
| Connection with People | 0.736 | 0.735 | 0.85 | 0.655 |
| Connection with Nature | 0.922 | 0.924 | 0.945 | 0.81 |
| Attitude towards Environment | 0.672 | 0.672 | 0.859 | 0.753 |
| Attitude towards Health | 0.804 | 0.804 | 0.872 | 0.63 |
| Family/Peer Environment and Value | 0.729 | 0.743 | 0.846 | 0.647 |
| Connection and Attachment to the Community | 0.796 | 0.799 | 0.88 | 0.71 |
| Culture | 0.78 | 0.78 | 0.901 | 0.82 |
| Local norm | 0.729 | 0.733 | 0.846 | 0.647 |
| ***Farmers' Market Advocates*** | | | | |
| Food Education and Learning Opportunities | 0.89 | 0.892 | 0.924 | 0.752 |
| Contribution to Environmental and Social Issues | 0.842 | 0.851 | 0.894 | 0.679 |
| Variety of Ingredients | 0.758 | 0.759 | 0.892 | 0.805 |
| Connection with People | 0.707 | 0.709 | 0.837 | 0.632 |
| Connection with Nature | 0.918 | 0.919 | 0.942 | 0.802 |
| Attitude towards Environment | 0.662 | 0.67 | 0.855 | 0.747 |
| Attitude towards Health | 0.823 | 0.825 | 0.883 | 0.654 |
| Family/Peer Environment and Value | 0.709 | 0.73 | 0.835 | 0.629 |
| Connection and Attachment to the Community | 0.784 | 0.792 | 0.873 | 0.697 |
| Culture | 0.726 | 0.73 | 0.879 | 0.785 |
| Local norm | 0.698 | 0.713 | 0.832 | 0.623 |
| ***Home Delivery Preferers*** | | | | |
| Food Education and Learning Opportunities | 0.853 | 0.856 | 0.901 | 0.695 |
| Contribution to Environmental and Social Issues | 0.827 | 0.832 | 0.885 | 0.659 |
| Variety of Ingredients | 0.792 | 0.793 | 0.906 | 0.828 |
| Connection with People | 0.643 | 0.644 | 0.807 | 0.583 |
| Connection with Nature | 0.885 | 0.888 | 0.921 | 0.744 |
| Attitude towards Environment | 0.72 | 0.721 | 0.877 | 0.781 |
| Attitude towards Health | 0.797 | 0.798 | 0.868 | 0.621 |
| Family/Peer Environment and Value | 0.629 | 0.639 | 0.799 | 0.571 |
| Connection and Attachment to the Community | 0.82 | 0.865 | 0.891 | 0.732 |
| Culture | 0.805 | 0.839 | 0.91 | 0.835 |
| Local norm | 0.705 | 0.718 | 0.834 | 0.627 |
| ***Sustainable Food Seekers*** | | | | |
| Food Education and Learning Opportunities | 0.87 | 0.872 | 0.911 | 0.719 |
| Contribution to Environmental and Social Issues | 0.836 | 0.841 | 0.891 | 0.671 |
| Variety of Ingredients | 0.803 | 0.804 | 0.91 | 0.835 |
| Connection with People | 0.714 | 0.73 | 0.839 | 0.635 |
| Connection with Nature | 0.9 | 0.901 | 0.93 | 0.769 |
| Attitude towards Environment | 0.712 | 0.72 | 0.874 | 0.776 |
| Attitude towards Health | 0.85 | 0.854 | 0.899 | 0.69 |
| Family/Peer Environment and Value | 0.756 | 0.761 | 0.86 | 0.671 |
| Connection and Attachment to the Community | 0.774 | 0.792 | 0.867 | 0.686 |
| Culture | 0.778 | 0.798 | 0.899 | 0.817 |
| Local norm | 0.771 | 0.775 | 0.867 | 0.685 |

Table A8: VIF for each hypothesis.

| | All-Sample | Segment1 | Segment2 | Segment3 | Segment4 | Segment5 |
|---|---|---|---|---|---|---|
| H1 Food Education and Learning Opportunities → Intention | 2.144 | 2.166 | 1.991 | 2.126 | 1.897 | 2.198 |
| H2 Contribution to Environmental and Social Issues → Intention | 2.469 | 2.205 | 2.374 | 2.236 | 2.073 | 2.716 |
| H3 Variety of Ingredients → Intention | 1.497 | 1.24 | 1.325 | 1.362 | 1.66 | 2.049 |
| H4 Connection with People → Intention | 1.855 | 1.563 | 1.782 | 1.696 | 1.821 | 2.123 |
| H5 Connection with Nature → Intention | 1.809 | 1.557 | 1.62 | 1.724 | 1.717 | 2.28 |
| H6 Attitude towards Environment → Food Education and Learning Opportunities | 1.643 | 1.496 | 1.436 | 1.739 | 1.552 | 2.026 |
| H7 Attitude towards Environment → Contribution to Environmental and Social Issues | 1.643 | 1.496 | 1.436 | 1.739 | 1.552 | 2.026 |
| H8 Attitude towards Environment → Variety of Ingredients | 1.643 | 1.496 | 1.436 | 1.739 | 1.552 | 2.026 |
| H9 Attitude towards Environment → Connection with People | 1.643 | 1.496 | 1.436 | 1.739 | 1.552 | 2.026 |
| H10 Attitude towards Environment → Connection with Nature | 1.643 | 1.496 | 1.436 | 1.739 | 1.552 | 2.026 |
| H11 Attitude towards Health → Food Education and Learning Opportunities | 1.643 | 1.496 | 1.436 | 1.739 | 1.552 | 2.026 |
| H12 Attitude towards Health → Contribution to Environmental and Social Issues | 1.643 | 1.496 | 1.436 | 1.739 | 1.552 | 2.026 |
| H13 Attitude towards Health → Variety of Ingredients | 1.643 | 1.496 | 1.436 | 1.739 | 1.552 | 2.026 |
| H14 Attitude towards Health → Connection with People | 1.643 | 1.496 | 1.436 | 1.739 | 1.552 | 2.026 |
| H15 Attitude towards Health → Connection with Nature | 1.643 | 1.496 | 1.436 | 1.739 | 1.552 | 2.026 |
| H26 Family/Peer Environment and Value → Attitude towards Environment | 1.38 | 1.194 | 1.358 | 1.241 | 1.35 | 1.615 |
| H27 Family/Peer Environment and Value → Attitude towards Health | 1.38 | 1.194 | 1.358 | 1.241 | 1.35 | 1.615 |
| H30 Connection and Attachment to the Community → Attitude towards Environment | 1.685 | 1.429 | 1.771 | 1.452 | 1.514 | 1.96 |
| H31 Connection and Attachment to the Community → Attitude towards Health | 1.685 | 1.429 | 1.771 | 1.452 | 1.514 | 1.96 |
| H34 Connection and Attachment to the Community → Family/Peer Environment and Value | 1.639 | 1.397 | 1.721 | 1.415 | 1.507 | 1.863 |
| H35 Culture → Attitude towards Environment | 1.323 | 1.307 | 1.415 | 1.214 | 1.261 | 1.348 |
| H36 Culture → Attitude towards Health | 1.323 | 1.307 | 1.415 | 1.214 | 1.261 | 1.348 |
| H39 Culture → Family/Peer Environment and Value | 1.307 | 1.307 | 1.405 | 1.167 | 1.182 | 1.329 |
| H40 Local Norm → Attitude towards Environment | 1.737 | 1.532 | 1.71 | 1.378 | 1.652 | 2.014 |
| H41 Local Norm → Attitude towards Health | 1.737 | 1.532 | 1.71 | 1.378 | 1.652 | 2.014 |
| H44 Local Norm → Family/Peer Environment and Value | 1.582 | 1.429 | 1.566 | 1.322 | 1.514 | 1.797 |

(Segment1: Conventional Shoppers, Segment2: Organic Enthusiasts, Segment3: Farmers' Market Advocates, Segment4: Home Delivery Preferers, Segment5: Sustainable Food Seekers)

620

Table A9: Discriminant validity

| ***Class***/Construct | Attitude towards Health | Connection and Attachment to the Community | Connection with Nature | Connection with People | Culture | Attitude towards Environment | Family/Peer Environment and Value | Food Education and Learning Opportunities | Contribution to Environmental and Social Issues | Intention | Local Norm | Variety of Ingredients |
|---|---|---|---|---|---|---|---|---|---|---|---|---|
| ***All-Sample*** | | | | | | | | | | | | |
| Attitude towards Health | 0.81 | | | | | | | | | | | |
| Connection and Attachment to the Community | 0.509 | 0.838 | | | | | | | | | | |
| Connection with Nature | 0.476 | 0.345 | 0.896 | | | | | | | | | |
| Connection with People | 0.524 | 0.505 | 0.505 | 0.8 | | | | | | | | |
| Culture | 0.26 | 0.446 | 0.234 | 0.289 | 0.904 | | | | | | | |
| Attitude towards Environment | 0.626 | 0.524 | 0.462 | 0.531 | 0.336 | 0.871 | | | | | | |
| Family/Peer Environment and Value | 0.474 | 0.425 | 0.302 | 0.483 | 0.328 | 0.492 | 0.803 | | | | | |
| Food Education and Learning Opportunities | 0.297 | 0.326 | 0.562 | 0.5 | 0.273 | 0.441 | 0.361 | 0.864 | | | | |
| Contribution to Environmental and Social Issues | 0.475 | 0.447 | 0.578 | 0.602 | 0.309 | 0.698 | 0.442 | 0.686 | 0.831 | | | |
| Intention | 0.249 | 0.266 | 0.409 | 0.368 | 0.224 | 0.405 | 0.283 | 0.587 | 0.537 | 1 | | |
| Local Norm | 0.61 | 0.582 | 0.405 | 0.508 | 0.413 | 0.632 | 0.485 | 0.343 | 0.528 | 0.341 | 0.809 | |
| Variety of Ingredients | 0.641 | 0.382 | 0.443 | 0.501 | 0.165 | 0.446 | 0.404 | 0.248 | 0.432 | 0.222 | 0.528 | 0.919 |
| ***Conventional Shoppers*** | | | | | | | | | | | | |
| Attitude towards Health | 0.797 | | | | | | | | | | | |
| Connection and Attachment to the Community | 0.316 | 0.839 | | | | | | | | | | |
| Connection with Nature | 0.371 | 0.284 | 0.917 | | | | | | | | | |
| Connection with People | 0.368 | 0.413 | 0.392 | 0.779 | | | | | | | | |
| Culture | 0.183 | 0.406 | 0.204 | 0.189 | 0.903 | | | | | | | |
| Attitude towards Environment | 0.576 | 0.374 | 0.401 | 0.41 | 0.252 | 0.848 | | | | | | |
| Family/Peer Environment and Value | 0.398 | 0.31 | 0.23 | 0.421 | 0.196 | 0.382 | 0.791 | | | | | |
| Food Education and Learning Opportunities | 0.313 | 0.312 | 0.552 | 0.476 | 0.223 | 0.419 | 0.389 | 0.864 | | | | |
| Contribution to Environmental and Social Issues | 0.401 | 0.37 | 0.508 | 0.534 | 0.254 | 0.67 | 0.375 | 0.689 | 0.82 | | | |
| Intention | 0.203 | 0.216 | 0.321 | 0.243 | 0.209 | 0.362 | 0.263 | 0.56 | 0.514 | 1 | | |
| Local Norm | 0.422 | 0.486 | 0.334 | 0.368 | 0.428 | 0.524 | 0.376 | 0.38 | 0.494 | 0.298 | 0.774 | |
| Variety of Ingredients | 0.558 | 0.245 | 0.315 | 0.396 | 0.113 | 0.36 | 0.364 | 0.276 | 0.331 | 0.18 | 0.39 | 0.916 |
| ***Organic Enthusiasts*** | | | | | | | | | | | | |
| Attitude towards Health | 0.793 | | | | | | | | | | | |
| Connection and Attachment to the Community | 0.497 | 0.843 | | | | | | | | | | |
| Connection with Nature | 0.436 | 0.319 | 0.9 | | | | | | | | | |
| Connection with People | 0.482 | 0.509 | 0.457 | 0.809 | | | | | | | | |
| Culture | 0.233 | 0.51 | 0.231 | 0.334 | 0.905 | | | | | | | |
| Attitude towards Environment | 0.551 | 0.526 | 0.409 | 0.532 | 0.366 | 0.868 | | | | | | |
| Family/Peer Environment and Value | 0.409 | 0.424 | 0.252 | 0.468 | 0.325 | 0.456 | 0.804 | | | | | |
| Food Education and Learning Opportunities | 0.23 | 0.294 | 0.506 | 0.489 | 0.328 | 0.424 | 0.332 | 0.865 | | | | |
| Contribution to Environmental and Social Issues | 0.439 | 0.447 | 0.55 | 0.604 | 0.339 | 0.731 | 0.41 | 0.66 | 0.838 | | | |
| Intention | 0.16 | 0.233 | 0.356 | 0.357 | 0.2 | 0.359 | 0.223 | 0.544 | 0.483 | 1 | | |
| Local Norm | 0.543 | 0.58 | 0.316 | 0.476 | 0.432 | 0.589 | 0.474 | 0.302 | 0.512 | 0.28 | 0.804 | |
| Variety of Ingredients | 0.604 | 0.317 | 0.35 | 0.412 | 0.15 | 0.341 | 0.333 | 0.137 | 0.347 | 0.125 | 0.439 | 0.927 |
| ***Farmers' Market Advocates*** | | | | | | | | | | | | |
| Attitude towards Health | 0.809 | | | | | | | | | | | |
| Connection and Attachment to the Community | 0.389 | 0.835 | | | | | | | | | | |
| Connection with Nature | 0.49 | 0.216 | 0.895 | | | | | | | | | |
| Connection with People | 0.534 | 0.368 | 0.49 | 0.795 | | | | | | | | |
| Culture | 0.151 | 0.364 | 0.127 | 0.185 | 0.886 | | | | | | | |
| Attitude towards Environment | 0.652 | 0.483 | 0.409 | 0.435 | 0.206 | 0.864 | | | | | | |
| Family/Peer Environment and Value | 0.436 | 0.345 | 0.194 | 0.415 | 0.314 | 0.416 | 0.793 | | | | | |
| Food Education and Learning Opportunities | 0.399 | 0.316 | 0.568 | 0.442 | 0.208 | 0.423 | 0.29 | 0.867 | | | | |
| Contribution to Environmental and Social Issues | 0.464 | 0.412 | 0.538 | 0.534 | 0.222 | 0.662 | 0.326 | 0.688 | 0.824 | | | |
| Intention | 0.307 | 0.268 | 0.368 | 0.249 | 0.179 | 0.371 | 0.141 | 0.531 | 0.417 | 1 | | |
| Local Norm | 0.58 | 0.484 | 0.29 | 0.461 | 0.267 | 0.591 | 0.347 | 0.302 | 0.458 | 0.271 | 0.789 | |
| Variety of Ingredients | 0.558 | 0.299 | 0.38 | 0.482 | 0.152 | 0.418 | 0.388 | 0.292 | 0.368 | 0.253 | 0.486 | 0.897 |
| ***Home Delivery Preferers*** | | | | | | | | | | | | |
| Attitude towards Health | 0.788 | | | | | | | | | | | |
| Connection and Attachment to the Community | 0.614 | 0.856 | | | | | | | | | | |
| Connection with Nature | 0.397 | 0.371 | 0.862 | | | | | | | | | |
| Connection with People | 0.522 | 0.508 | 0.483 | 0.763 | | | | | | | | |
| Culture | 0.304 | 0.344 | 0.167 | 0.308 | 0.914 | | | | | | | |
| Attitude towards Environment | 0.597 | 0.502 | 0.464 | 0.542 | 0.269 | 0.884 | | | | | | |
| Family/Peer Environment and Value | 0.488 | 0.336 | 0.228 | 0.391 | 0.379 | 0.416 | 0.756 | | | | | |
| Food Education and Learning Opportunities | 0.218 | 0.267 | 0.536 | 0.388 | 0.124 | 0.45 | 0.195 | 0.834 | | | | |
| Contribution to Environmental and Social Issues | 0.484 | 0.373 | 0.509 | 0.545 | 0.223 | 0.671 | 0.359 | 0.591 | 0.812 | | | |
| Intention | 0.321 | 0.301 | 0.444 | 0.328 | 0.179 | 0.553 | 0.317 | 0.569 | 0.538 | 1 | | |
| Local Norm | 0.583 | 0.558 | 0.405 | 0.491 | 0.349 | 0.622 | 0.446 | 0.266 | 0.511 | 0.413 | 0.792 | |
| Variety of Ingredients | 0.678 | 0.474 | 0.393 | 0.551 | 0.163 | 0.513 | 0.382 | 0.138 | 0.446 | 0.262 | 0.534 | 0.91 |
| ***Sustainable Food Seekers*** | | | | | | | | | | | | |
| Attitude towards Health | 0.831 | | | | | | | | | | | |
| Connection and Attachment to the Community | 0.591 | 0.828 | | | | | | | | | | |
| Connection with Nature | 0.583 | 0.423 | 0.877 | | | | | | | | | |
| Connection with People | 0.622 | 0.588 | 0.596 | 0.797 | | | | | | | | |
| Culture | 0.353 | 0.466 | 0.281 | 0.305 | 0.904 | | | | | | | |
| Attitude towards Environment | 0.712 | 0.605 | 0.542 | 0.592 | 0.408 | 0.881 | | | | | | |
| Family/Peer Environment and Value | 0.564 | 0.533 | 0.403 | 0.542 | 0.381 | 0.613 | 0.819 | | | | | |
| Food Education and Learning Opportunities | 0.341 | 0.399 | 0.577 | 0.542 | 0.278 | 0.449 | 0.398 | 0.848 | | | | |
| Contribution to Environmental and Social Issues | 0.535 | 0.523 | 0.639 | 0.637 | 0.342 | 0.684 | 0.531 | 0.697 | 0.819 | | | |
| Intention | 0.312 | 0.318 | 0.423 | 0.403 | 0.247 | 0.406 | 0.32 | 0.569 | 0.559 | 1 | | |
| Local Norm | 0.746 | 0.649 | 0.524 | 0.594 | 0.435 | 0.72 | 0.572 | 0.345 | 0.545 | 0.351 | 0.828 | |
| Variety of Ingredients | 0.69 | 0.472 | 0.625 | 0.607 | 0.204 | 0.549 | 0.478 | 0.355 | 0.56 | 0.292 | 0.646 | 0.914 |

**Interclass comparisons**

Consumer segments were compared in terms of demographic characteristics. For continuous scale responses, one-way analysis of variance (ANOVA) and Games-Howell multiple comparisons were used. For ordinal scale responses, Kruskal-Wallis and Dunn-Bonferroni multiple comparisons were employed. The analysis was conducted using IBM SPSS Statistics (ver.29.0.2.0).

**Interclass comparisons from sociodemographic perspective**

Analysis of sociodemographic characteristics across the identified consumer segments revealed significant differences in age (Mean: 50.7, SD: 13.6, $F = 32.6$, $p < 0.01$), annual household income ($H = 70.4$, $p < 0.01$), and annual individual income ($H = 73.1$, $p < 0.01$). "Sustainable Food Seekers" demonstrated the youngest mean age, with statistically significant differences ($p < 0.05$) observed between this group and the "Conventional Shoppers," "Organic Enthusiasts," and "Home Delivery Preferers" segments. Regarding household income, "Organic Enthusiasts" were predominantly characterized by incomes below 4 million yen, showing significant differences from all other segments. Similarly, in terms of annual individual income levels, "Sustainable Food Seekers" showed significant differences from "Conventional Shoppers," "Organic Enthusiasts" segments, with 4 to 6 million JPY and Over 6 million JPY each accounting for around 20%. These results are shown in Table A10 and A11.

These sociodemographic insights highlight the diverse nature of consumer groups within the market, emphasizing the unique characteristics of each segment.

Table A10: Results of One-way ANOVA and Games-Howell's Post Hoc Multiple Comparison for sociodemographic

| | All-Sample (*N* = 2484) | | Segment1 (*N* = 460, 18.5%) | | Segment2 (*N* = 839, 33.8%) | | Segment3 (*N* = 280, 11.3%) | | Segment4 (*N* = 250, 10.0%) | | Segment5 (*N* = 655, 26.4％) | | *F* | *p* |
|---|---|---|---|---|---|---|---|---|---|---|---|---|---|---|
| | Mean | SD | Mean | SD | Mean | SD | Mean | SD | Mean | SD | Mean | SD | | |
| Age | 50.69 | 13.60 | 51.14a | 12.74 | 54.11b | 13.08 | 48.64a, c | 13.45 | 51.59a, b | 14.82 | 46.53c | 13.18 | 32.6 | < 0.001 |

(Segment1: Conventional Shoppers, Segment2: Organic Enthusiasts, Segment3: Farmers' Market Advocates, Segment4: Home Delivery Preferers, Segment5: Sustainable Food Seekers)
Note. Superscripts a, b and c indicate statistically significant difference exist between the groups ($p < .05$).

Table A11: Results of Kruskal Wallis and Dunn-Bonferroni Multiple Comparisons for annual income

| Variable | Segment | *N* | *df* | *H* | *p* | Significant Difference by Dunn-Bonferroni test |
|---|---|---|---|---|---|---|
| Annual household income | 1 | 460 | 4 | 70.4 | < 0.001 | 2-1, 2-3, 2-4, 2-5 |
| | 2 | 839 | | | | |
| | 3 | 280 | | | | |
| | 4 | 250 | | | | |
| | 5 | 655 | | | | |
| Annual personal income | 1 | 460 | 4 | 73.1 | < 0.001 | 1-3, 1-4, 1-5, 2-3, 2-4, 2-5 |
| | 2 | 839 | | | | |
| | 3 | 280 | | | | |
| | 4 | 250 | | | | |
| | 5 | 655 | | | | |

(Segment1: Conventional Shoppers, Segment2: Organic Enthusiasts, Segment3: Farmers' Market Advocates, Segment4: Home Delivery Preferers, Segment5: Sustainable Food Seekers)